\documentclass[aps,pre,twocolumn,showpacs]{revtex4}
\usepackage{latexsym}
\usepackage{amsmath,amssymb,epsfig,graphicx}
\begin{document}
\title{The noisy Hegselmann-Krause model for opinion dynamics}
\author{Miguel Pineda}
\email[]{mpineda@fis.usb.ve}
\affiliation{Department of Physics, Universidad Sim\'on Bol\'{\i}var, Caracas 1080, Venezuela}
\author{Ra\'ul Toral and Emilio Hern\'andez-Garc\'{i}a\ }
\affiliation{IFISC, Instituto de F\'{i}sica Interdisciplinar y Sistemas Complejos (CSIC-UIB), 07122 Palma de Mallorca, Spain}
\date{\today}

\begin{abstract}
In the model for continuous opinion dynamics introduced by
Hegselmann and Krause, each individual moves to the average
opinion of all individuals within an area of confidence. In
this work we study the effects of noise in this system. With
certain probability, individuals are given the opportunity to
change spontaneously their opinion to another one selected
randomly inside the opinion space with different rules. If the
random jump does not occur, individuals interact through the
Hegselmann-Krause's rule. We analyze two cases, one where
individuals can carry out opinion random jumps inside the whole
opinion space, and other where they are allowed to perform
jumps just inside a small interval centered around the current
opinion. We found that these opinion random jumps change the
model behavior inducing interesting phenomena. Using pattern
formation techniques, we obtain approximate analytical results
for critical conditions of opinion cluster formation. Finally,
we compare the results of this work with the noisy version of
the Deffuant et al. model for continuous-opinion dynamics.
\end{abstract}
\maketitle
\section{Introduction}
\label{intro}
In a social system, the opinion of the individuals determines
the character of their mutual interactions. But at the same
time, the formation and subsequent evolution of people's
opinion are complex phenomena affected by affinities and
contracts between the members of the society. This complex
behavior is specially observed in situations when a common
decision needs to be taken by the individuals. During such a
cooperative task it usually happens that either a single
position emerges or the population evolves to a state of
coexistence of different opinions. It is natural to talk about
those processes within the framework of interacting particles,
this being one of the reasons why nowadays many physicists
address the study of opinion formation in large groups using
ideas borrowed from statistical physics and non-linear science
\cite{stauffer1,castellano}. The introduction of new
information-communication technologies and the availability of
large data sets have also contributed to develop this
interdisciplinary research field.

In recent years, two models where the opinion of an individual
can vary continuously have raised the interest of the
scientific community \cite{lorenz1,lorenz2}. Such continuous
models have been introduced independently by Deffuant and
collaborators (DW Model) \cite{deffuant1} and Hegselmann and
Krause (HK model) \cite{krause1,krause2,fortunato,slanina}. The
two models implement the so-called bounded confidence mechanism
by which two individuals only influence each other if their
opinions differ less than some given amount
\cite{granovetter1,axelrod1}. Another common important
ingredient of both models is an agreement mechanism, by which 
individuals that satisfy the bounded confidence condition 
adjust their opinions towards an average value. The fundamental 
difference between the models is
materialized in the definition of who communicates with whom at
once \cite{urbin1}. In the DW model, two randomly chosen
individuals meet and a pairwise averaging is implemented, while
there is an extra parameter that controls how fast the opinions
converge \cite{laguna1,porfiri1}. This model is suitable to
describe situations where individuals meet in small groups and
exchange information face-to-face. In the HK model, the
communication takes place in large groups and individuals move
their own opinions to the average opinion of all individuals
which lie in the area of confidence.

Although one expects considerable differences between the two
models when the number of individuals is large, it has been
well established that they always lead to a final state in
which either perfect consensus is reached or the population
splits into a set of opinion clusters each of them holding
exactly the same opinion \cite{lorenz1,redner1}. However, in
real social systems, public opinion does not reach such ideal
states of complete consensus. In this regard and with the aim
to make models of continuous opinion dynamics more realistic,
recent works have introduced additional elements of randomness
to the DW model. This new ingredient has been interpreted as a
``self-thinking'' or ``free-will'', where individuals change
their opinion in a random way \cite{pineda1,pineda2}, as the
death of an individual and the birth of a new one
\cite{jensen1,carletti1}, or simply, as the replacement of
individuals by new ones in systems where the total size is not
fixed \cite{maxi1}.

Nowadays, with the introduction of new
information-communication technologies, an effective global
exchange of information in large groups is easily achieved. In
this sense, we believe that the HK model deserves more
attention, particularly when a sort of randomness is added to
the original rules \cite{igor}. Following this motivation, in this paper we
generalize the HK dynamical rules to incorporate additional
random elements, or ``noise". Our aim is to analyze which
aspects of the original dynamics are robust against noise and
which additional complex collective phenomena can emerge as a
result. In our generalization, individuals are allowed to
change spontaneously their opinion with certain probability
\cite{pineda1,pineda2}. If this random jump does not occur,
individuals can then perform interactions through the HK's
rules. We analyze two cases of noise that have been already
successfully implemented in the DW model
\cite{pineda1,pineda2}: In the first case, individuals are
allowed to perform opinion random jumps to any point in the
full opinion space, while in the second case, individuals can
perform a random jump in their opinion to a new value located
inside a small interval centered around the current opinion. We
show that these new ingredients are able to induce novel
phenomena in the HK model. In both cases, we have found an
order-disorder transition above a critical value of the noise
intensity. In the disordered state the opinion distribution
tends to be uniform, while for the ordered state, a set of
noisy opinion clusters are formed. Using a linear stability
analysis we derive approximate conditions for the stability of
noisy opinion clusters. Our analytical results are in
qualitative agreement with Monte Carlo simulations.

The next section presents the HK model in the presence of
noise. Section 3 contains extensive results on the model
behavior obtained by Monte Carlo simulations. The
order-disorder transition is analyzed through a linear
stability analysis in Section 4. Section 5 is devoted to
compare the noisy HK model with the noisy DW model. Conclusions
are presented in Section 6.
\section{The noisy Hegselmann-Krause model}
\label{sec:1}
The original HK model was introduced as a nonlinear extension
of previous models of social influence
\cite{krause1,krause2,fiedrick1}. In this section, we consider
a modification of the model in which noise is added to the
original HK rules, resulting in a random change of an
individual's opinion. To begin the analysis, let us consider a
system composed by $N$ individuals ($i=1,...,N$). At (discrete)
time $n$ each individual $i$ is endowed with a continuous
opinion $x^{n}_{i}$, taking values in a continuous
one-dimensional interval $x^{n}_{i}\in [0,L]$, where $L$ is the
range of opinion space. At time-step $n$ a randomly chosen
individual $i$ has a probability $m$ of spontaneously changing
his opinion to a new random value, and a probability $1-m$ to
move to the average opinion of all individuals (including
himself) which lie in his interval of confidence of width
$2\epsilon$. The case $m=0$ corresponds to the standard HK
model, in which the opinion of the individual $i$, at the next
step $n+1$, is given by
\begin{equation}
x^{n+1}_{i}=\frac{\sum\limits_{j:|x^{n}_{i}-x^{n}_{j}| \leq
\epsilon}x^{n}_{j}}{|\{j:|x^{n}_{i}-x^{n}_{j}| \leq \epsilon\}| } \ ,
\label{eq:hkrule}
\end{equation}
where the sum is over the individuals $j$ whose opinions differ
from $x^{n}_{i}$ by at most $\epsilon$, and
$|\{j:|x^{n}_{i}-x^{n}_{j}| \leq \epsilon\}|$ is the number of
such individuals. The procedure is repeated by selecting at random another
individual and so on \cite{fortunato,fortunato1}. The parameter
$\epsilon$, which runs from $0$ to $L$, is the confidence
parameter. We introduce the time variable $t=n/N$ measuring the
number of Monte Carlo steps (MCS), or the number of opinion
updates per individual. 

As far as the range  of the random jumps (the maximum interval in which individuals can change
spontaneously their opinions) is concerned, we distinguish two
simple scenarios:

\begin{itemize}
\item[(1)]Unlimited random jumps to any point inside the interval
$[0,L]$, meaning that the new opinion $x_i^{n+1}$ can take any value in the whole opinion
space $[0,L]$ \cite{pineda1}.
\item[(2)]Bounded random jumps inside the interval $[-\gamma,\gamma]$, with
$\gamma \leq L$. i.e.  the new opinion $x_i^{n+1}$ will lie in the interval
$(x_i^{n}-\gamma,x_i^{n}+\gamma)$ \cite{pineda2}.
\end{itemize}

In both scenarios, the new random value is adopted uniformly
within the allowed interval. In the second case, it is possible
that opinions leave the bounded opinion space $[0,L]$. To avoid
this problem, we will consider {\it adsorbing boundary
conditions} in which opinions that try to go away towards the
left or towards the right of the interval $[0,L]$ are set to
$0$ and $L$, respectively. The more convenient from the
mathematical point of view {\it periodic boundary conditions},
where the opinion space $[0,L]$ is considered to be wrapped on
a circle, will be also considered in particular cases as
properly mentioned. For each particular case, the type of final
configurations reached by the system will depend on the values
of the threshold $\epsilon$, the noise intensity $m$, and/or
the parameter $\gamma$. Although we will keep the notation $L$
when referring to the range of opinion space, all the results
of this paper are for $L=1$. Results for other $L$ values can
be easily translated from ours by making the rescaling
$\epsilon\to\epsilon/L$ and $\gamma\to\gamma/L$.

This noisy HK model can be described in terms of an approximate
density-based master equation for the probability density
$P(x,t)$ that an individual holds opinion $x$ at time $t$. This
equation can be written as
\begin{equation}
\begin{split}
\frac{\partial P(x,t)}{\partial t} &=
(1-m)\left[\int_L  dx_1 P(x_1,t)\left( \delta( x-\langle x\rangle_{x_1})-\delta(x-x_1)\right)\right]\\
                                   &  +m\left[G(x,t)-P(x,t)\right],
\label{hkme}
\end{split}
\end{equation}
where $\langle x\rangle_{x_1}$ is the average position of the individuals
within distance $\epsilon$ of opinion $x_1$, i.e.
\begin{equation}
 \langle x\rangle_{x_1} =\frac{\int_{x_1-\epsilon}^{x_1+\epsilon} u P(u,t) du}{\int_{x_1-\epsilon}^{x_1+\epsilon} P(u,t) du}.
\label{eq:hkmean}
\end{equation}
In this average, the denominator is the normalization by the
probability mass in the interval $[x_1-\epsilon,x_1+\epsilon]$
while the numerator is the first moment in that interval. In
Eq. (\ref{hkme}) the term proportional to $m$ describes the
random jumps, whereas the one proportional to $(1-m)$
represents the original HK rules. For unlimited random jumps,
the function $G(x,t)$ is the homogeneous distribution
$P_{h}(x,t)=1/L$ \cite{pineda1}, whereas for bounded random
jumps with adsorbing boundary conditions \cite{pineda2},
\begin{equation}
G(x,t)=
\left\{ \begin{array}{ll}
 \delta(x) \int_0^{\gamma} dx^{'}\frac{\gamma-x^{'}}{2\gamma}P(x^{'},t)\\
+\int_0^{x+\gamma} \frac{dx^{'}}{2\gamma}P(x^{'},t), & \mbox{if } x\leq\gamma, \\
\\
\int_{x-\gamma}^{x+\gamma} \frac{dx^{'}}{2\gamma}P(x^{'},t), & \mbox{if } \gamma\leq x \leq L-\gamma,\\
\\
\delta(x-L) \int_{L-\gamma}^{L}
dx^{'}\frac{-L+\gamma+x^{'}}{2\gamma}P(x^{'},t)\\
+\int_{x-\gamma}^{L} \frac{dx^{'}}{2\gamma}P(x^{'},t), & \mbox{if } x \geq L-\gamma.
\end{array} \right.
\label{G}
\end{equation}

Before we continue  with the analysis, let us summarize some of
the most relevant features observed in the original noiseless
HK model ($m=0$) \cite{fortunato1}. Eq.~(\ref{hkme}) with $m=0$
provides a mean-field description (in the sense that
correlations between agents' opinions have been neglected) of
the process of selecting a random individual and changing his
opinion to the average of the individuals in a neighborhood of
size $2\epsilon$. Starting from uniformly distributed random
opinions, Monte Carlo simulations show that for $\epsilon>0$
the system either reaches a final state of complete consensus
or splits into a number of opinion clusters separated by a
distance larger than $\epsilon$. In the case of $L=1$ and
uniform initial distribution of opinions, $P(x,t=0)=1$ for $x
\in [0,1]$ and $P(x,t=0)=0$ otherwise, the result given by the 
master equation is that for $\epsilon\geq 0.19$ only a big cluster emerges and the steady
state distribution is $P_{\infty}(x)=\lim_{t\rightarrow\infty}P(x,t)=\delta(x-1/2)$,
whereas for smaller values of $\epsilon$ a series of
bifurcations and nucleation of clusters occur. In this
clustering regime it is found that
$P_{\infty}(x)=\sum^{n_{c}}_{i=1}m_{i}\delta(x-x_{i})$ with
$|x_{i}-x_{j}|>\epsilon$ for all $i\neq j$ and
$\sum^{n_{c}}_{i=1}m_{i}=1$, where $n_{c}$ is the number of opinion
clusters, $x_{i}$ is the position of a cluster and $m_{i}$ its
mass. Unlike other bounded confidence models, the noiseless HK
model evolving from uniform initial conditions does not exhibit
the so-called minor or low-populated clusters at the extreme
and between high populated clusters \cite{lorenz1,fortunato1}.
These minor cluster can appear when starting from more
asymmetric initial conditions.

\section{Monte Carlo simulations}
\label{sec:2}
It is a well-known fact that in continuous opinion dynamics the
master equation and the Monte Carlo simulations do not always
agree due to finite-size induced fluctuations and to having
neglected the correlations between agents. In this section, we
present the main phenomena obtained from Monte Carlo
simulations with a finite system of $N$ individuals and initial
conditions  randomly and uniformly distributed in the opinion
space interval $[0,L]$.

\subsection{Unlimited random jumps}
In this subsection, we will analyze the impact of unlimited
opinion random jumps on the original HK model. As it was
mentioned above, a randomly chosen individual can change with
probability $m$ his opinion to a random opinion inside the full
interval $[0,L]$. Otherwise, with probability $1-m$ the
individual interacts with their compatible neighbors following
the HK's rule. We will show that the interplay between the
confidence parameter $\epsilon$ and the noise intensity $m$
induces very interesting phenomena.
\begin{figure}[ht]
\begin{center}
\mbox{\includegraphics[clip,angle=270,width=.3 \textwidth]{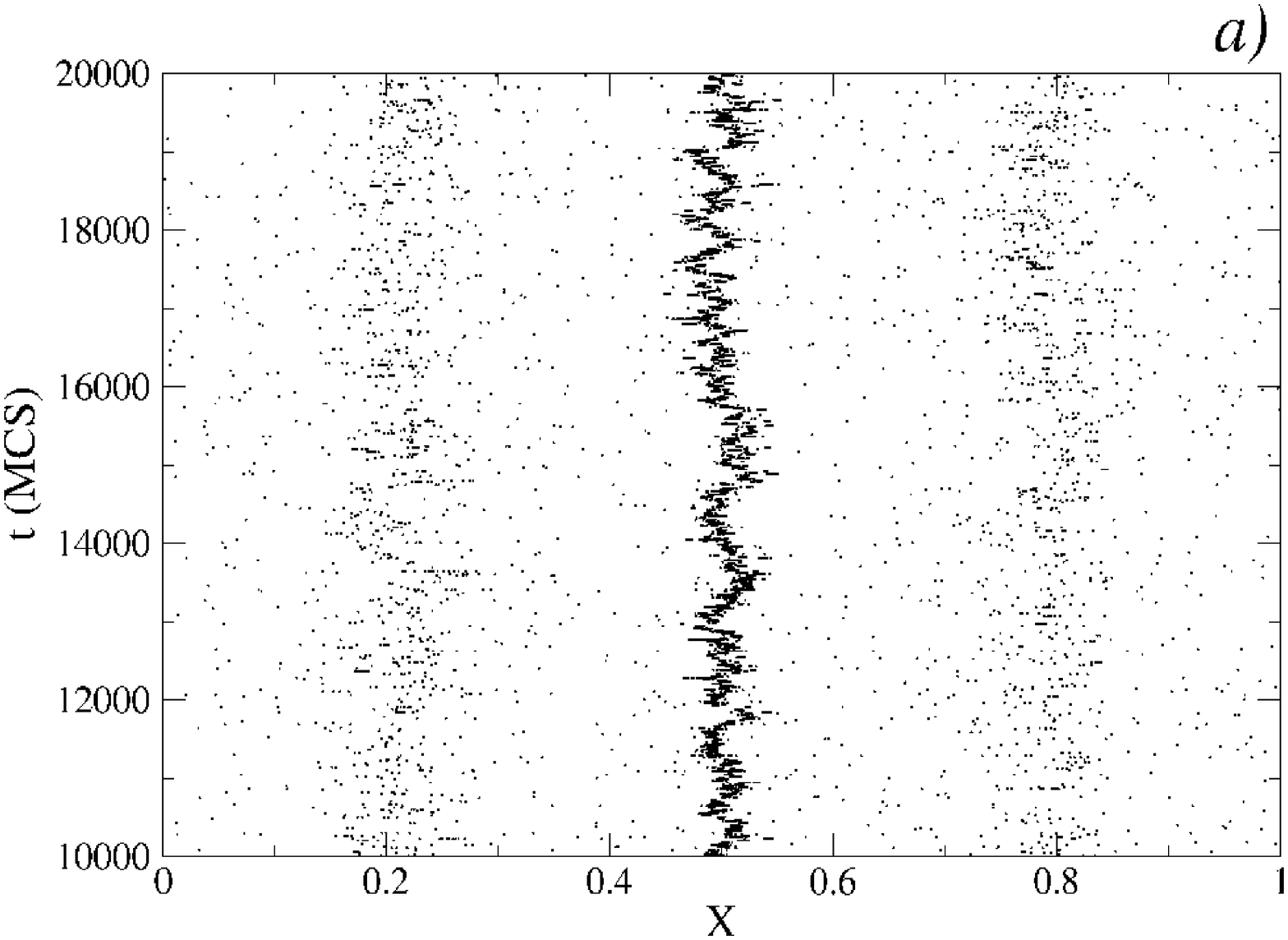}}
\end{center}
\begin{center}
\mbox{\includegraphics[clip,angle=270,width=.3 \textwidth]{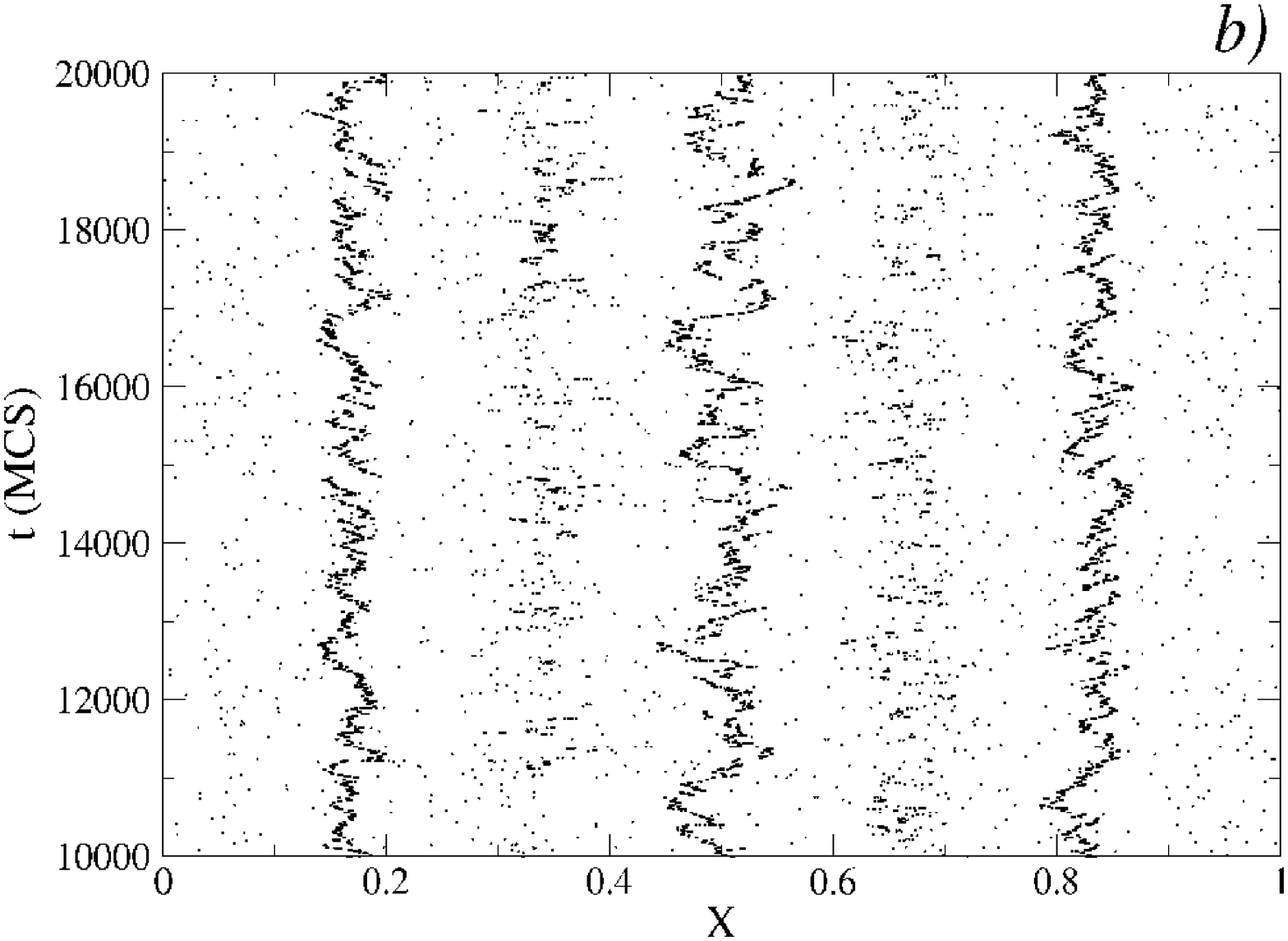}}
\end{center}
\caption{\label{fig1} Time series in opinion space for $m=0.02$. a) The case $\epsilon=0.27$, where
only one big cluster is formed when $m=0$. b) The case $\epsilon=0.127$, where only three big
clusters are formed when $m=0$. The number of individuals is $N=1000$, but only $100$ randomly chosen among
them are plotted to avoid saturation of the plot. Note the formation of low-populated opinion clusters
at the extremes and between high populated clusters when $m>0$. The opinion space runs from $0$ to $L=1$.
The initial condition at $t=0$ was uniform in $[0,1]$ and data starts to be plotted after long
enough simulation time.}
\end{figure}

One of the most distinct features of the noiseless HK model
\cite{lorenz1,fortunato1} is the lack of low-populated opinion
clusters at the extremes and between high populated clusters
when the initial condition is uniform in opinion space. The
absence of this class of minor clusters, which are typically
observed in other models of continuous opinion dynamics, is a
consequence of the fully connected and mutual convergence of
all the individuals since the very beginning. In other
continuous opinion dynamics systems, like the DW model, the
interaction is between randomly chosen pairs of individuals and
therefore some opinions are not able to interact enough times
to enter the basin of attraction of the big clusters.
Nevertheless, when noise is introduced, one notices in the HK
model the appearance of low-populated clusters for certain
values of $\epsilon$. For example, Fig.~\ref{fig1} shows time series 
of the opinions from Monte Carlo simulations for values of 
$\epsilon$ such that only one (panel a) or three (panel b) 
clusters are formed when $m=0$. Figure~\ref{fig1}(a) shows that for $m>0$ a
pattern of three opinion clusters is established. The two
extreme clusters are low populated and the central one is
composed by the vast majority of agents. Figure~\ref{fig1}(b)
shows a similar case but for a lower value of $\epsilon$. In
this case, it is clear that low-populated opinion clusters also
appear between clusters with higher populations. Under this
type of noise the whole opinion space can be covered and
therefore low-populated clusters have more chance to be
established out of the range of interaction of highly populated
clusters. In fact, they start to increase their population when
increasing the noise intensity $m$.
\begin{figure}[ht]
\begin{center}
\mbox{\includegraphics[clip,angle=270,width=.25 \textwidth]{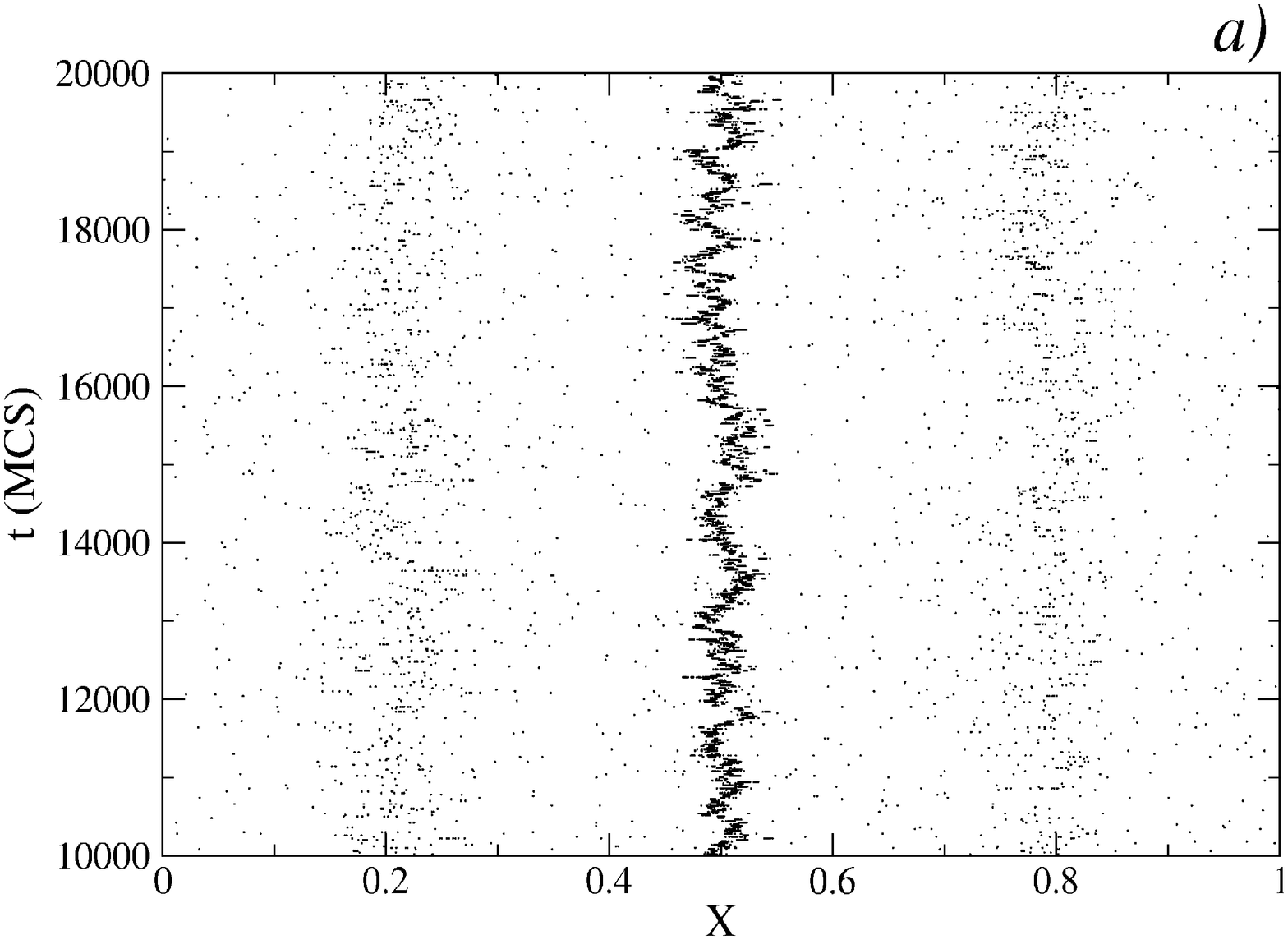}}
\end{center}
\begin{center}
\mbox{\includegraphics[clip,angle=270,width=.25 \textwidth]{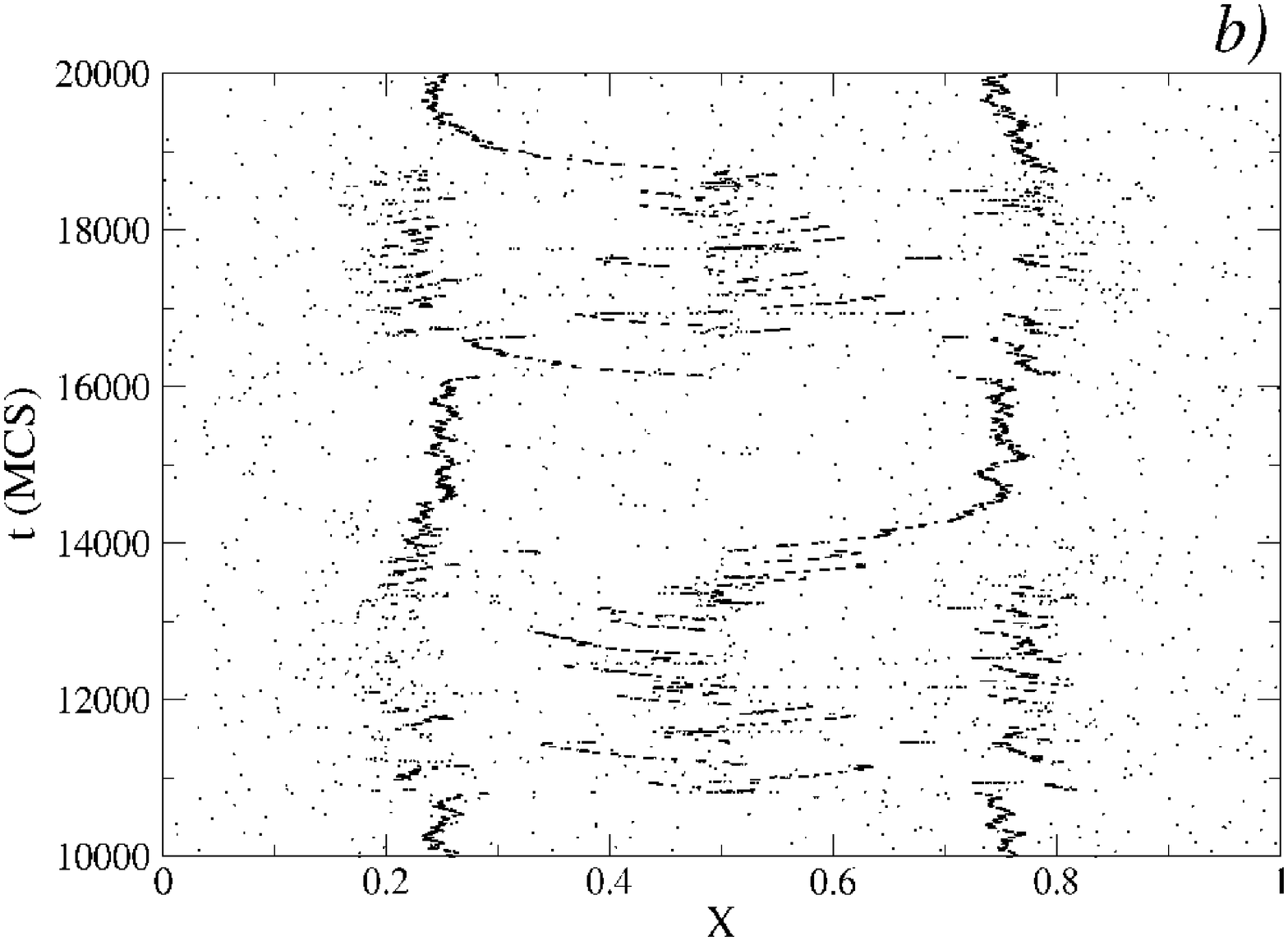}}
\end{center}
\begin{center}
\mbox{\includegraphics[clip,angle=270,width=.25 \textwidth]{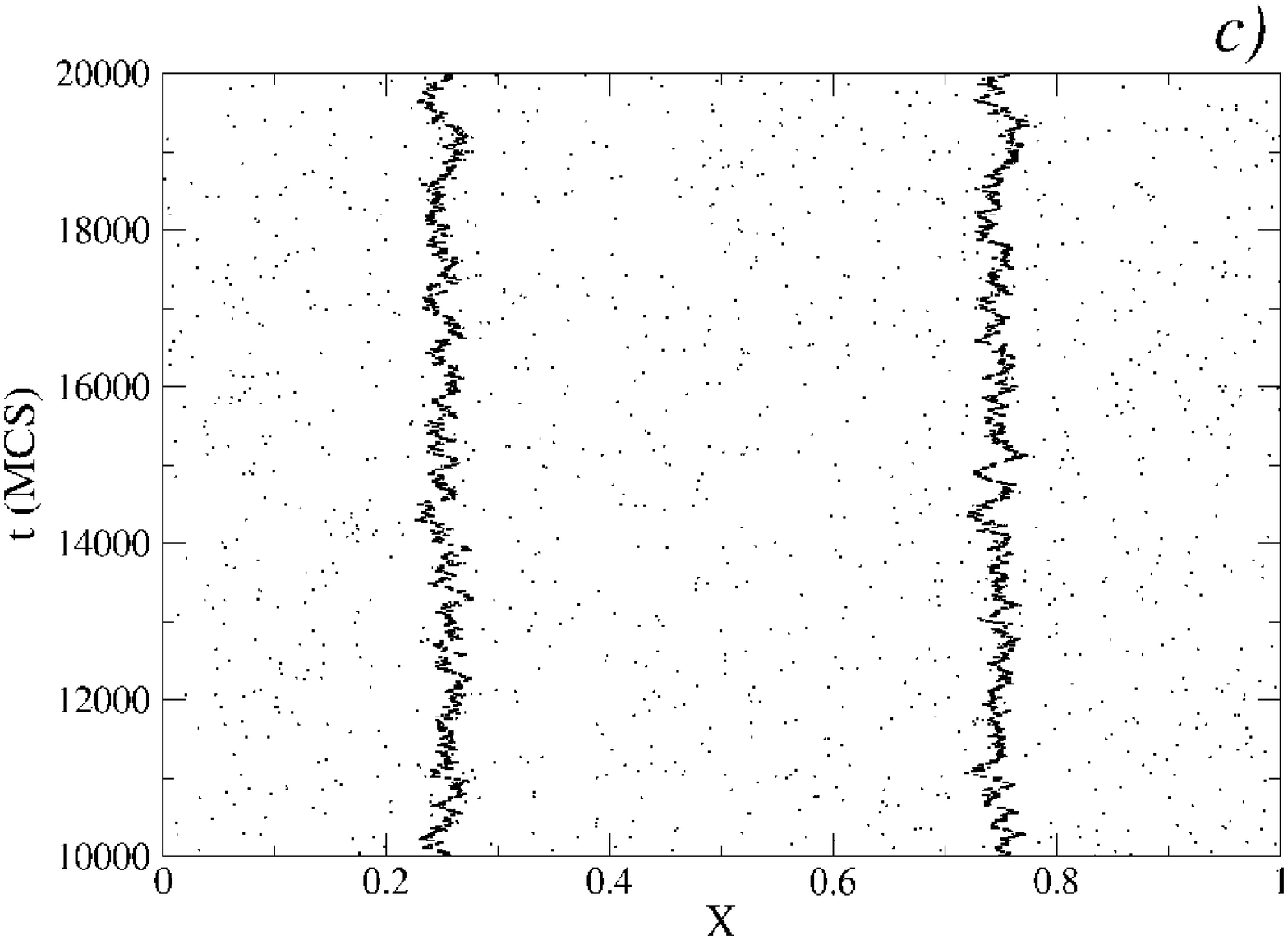}}
\end{center}
\caption{\label{fig2} Time series in opinion space for unlimited jumps at three
values of  $\epsilon$ for $m=0.02$ and $N=1000$ (only $100$ agents are plotted to avoid
saturation of the plot). At $\epsilon=0.270$ (panel a) a single high populated cluster
dominates over two lateral low-populated clusters. At $\epsilon=0.23$0 (panel c) two polarized
opinion cluster appear.  At $\epsilon=0.242$ (panel b) the system
randomly jumps between these two states.  The panels represent values of the confidence
parameter $\epsilon$ for which the noiseless HK model ($m=0.0$) is near a transition from one big
cluster to two big ones. Note also in (a) the formation of low-populated extreme opinion
clusters that play an important role for jumps. The opinion space runs from
$0$ to $L=1$ and data starts to be plotted after long enough simulation time.}
\end{figure}

Similarly to \cite{pineda1}, we report a bistable behavior for
narrow bands of $\epsilon$ near the bifurcation transitions
between one stable configuration and the next one. As is
typical in bistable situations, we observe that the inherent
fluctuations of a finite-size system induce transitions between
one state and back. These jumps are, for instance, observed in
Monte Carlo simulations for $\epsilon=0.242$ near the
transition for one big cluster to two big ones.
Figure~\ref{fig2}(b) shows several jumps between both states.
Also note that low-populated clusters always exist and play a
key role in the transitions [see Figs.~\ref{fig2}(b) and
~\ref{fig2}(c)].

\subsection{Bounded random jumps}
\begin{figure}[ht]
\begin{center}
\mbox{\includegraphics[clip,angle=270,width=.3 \textwidth]{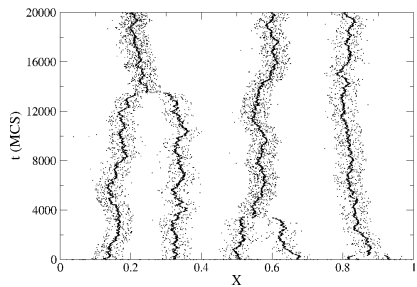}}
\end{center}
\caption{\label{fig3} Time series of the opinion distribution for bounded
jumps with $\epsilon=0.05$, $\gamma=0.04$, and $m=0.05$. Opinions form clusters that execute
random walks, and successive merging of
clusters occurs after collision. At very long time (not shown) only one big cluster of finite
width remains. In this simulation adsorbing boundary conditions
are considered. The opinion space runs from $0$ to $L=1$ and only $100$
opinions are plotted out of $N=1000$.}
\end{figure}

We now allow individuals to perform, with probability $m$,
jumps limited to the interval $[-\gamma,\gamma]$ centered
around their current opinion. We found that, when adsorbing
boundary conditions are considered, noisy opinion clusters still 
form for small and moderate noise intensity $m$.
However, for $\gamma$ small the clusters do not form symmetric
patterns around the mean opinion $0.5$. Instead, the centers of
mass of each one of them perform a random walk along the whole
opinion space until eventually they collide to form only one
big opinion cluster. Figure~\ref{fig3} shows the successive
merging of clusters occurring after collisions

\begin{figure}[ht]
\begin{center}
\mbox{\includegraphics[clip,angle=270,width=.3 \textwidth]{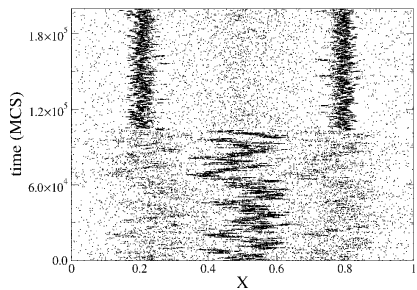}}
\end{center}
\caption{\label{fig4} Time series of the opinion distribution for bounded
jumps with $\epsilon=0.252$, $\gamma=0.495$, and $m=0.07$. Note the transition
from one big cluster with two sidebands to a state of only two big clusters.
We were unable to find transitions back to one single cluster. The opinion space
runs from $0$ to $L=1$ and only $100$ opinions are plotted out of $N=1000$.}
\end{figure}

For large values of $\gamma$, a stable pattern of opinion
clusters with a reduction of their wandering is observed. Under
these conditions, one can also find regions of bistability
where the inherent fluctuations of our finite system take the
system from one state to another. For the case presented in
Fig.~\ref{fig4}, transitions back were not found even for very
long simulation times. The figure just shows an early jump from
a state of a big opinion cluster and two smaller ones to a
state of two big opinion clusters.
\section{Order-disorder transitions}
\label{sec:3}
In many systems, one of the main effects of noise is to induce
an order-disorder transition. In this sense, opinion dynamics
is not the exception
\cite{pineda1,pineda2,jensen1,weron1,ben1}. In general we
expect that when in our noisy model the intensity $m$ is larger
than a critical value $m_{c}$, the patterns of opinion would
become blurred such that the corresponding maxima of the
distributions $P(x,t)$ are not evident, implying the
destruction of opinion clusters and the establishment of a
highly homogeneous state far from the boundaries. This effect
can be analyzed using Monte Carlo simulations or the
corresponding density-based master equation. We now present a
linear stability analysis of the master equation in order to
obtain analytical conditions for the existence of opinion
clusters under noise. In particular, the linear stability
analysis of the unstructured solution of Eq.~(\ref{hkme}) is
performed. Then, the obtained expressions are compared with
Monte Carlo simulations.
\begin{figure}[ht]
\begin{center}
\mbox{\includegraphics[clip,angle=270,width=.3 \textwidth]{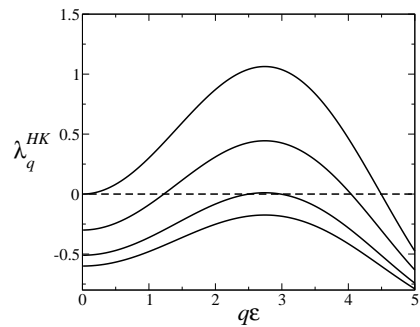}}
\end{center}
\caption{\label{fig5} This figure shows the growth rate, $\lambda^\textrm{HK}_{q}$,
for the case of unlimited random jumps with noise intensities
$m=0.0$, $0.3$, $0.51$, $0.6$, from top to bottom. Its shows that the growth rate
becomes negative for $m>m_{c} \approx 0.51$.}
\end{figure}
If one neglects the influence of the borders or assumes that
the opinion space is wrapped on a circle, the steady solution
$P_{h}(x)=1/L$ is an approximation to the unstructured steady
solution of Eq.~\ref{hkme}. It allows us to introduce
$P(x,t)=1/L+A_{q}\exp({iqx+\lambda^\textrm{HK}_{q}t})$, where
$\lambda^\textrm{HK}_{q}$ represents the growth rate of
periodic perturbations, $q$ is the corresponding wavenumber,
and $A_{q}$ the amplitude. After introducing this ansatz in
Eq.~(\ref{hkme}) we find the growth rate of the mode $q$:
\begin{equation}
\lambda^\textrm{HK}_{q}=(1-m)\left[\frac{\sin(q\epsilon)}{q\epsilon}-\cos(q\epsilon)\right]+mH(q).
\label{eq:drgeneral}
\end{equation}
The function $H(q)$ is equal to $-1$ for the case of unlimited
random jumps inside the whole opinion space. For bounded random
jumps inside the interval
$[x^{n}_{i}-\gamma,x^{n}_{i}+\gamma]$, we consider the case of
small values of $\gamma$ because in this case the boundary
effects become less important and the linear stability analysis
of the homogeneous state $P_h=1/L$ becomes valid. In this
situation, the second case of Eq.~(\ref{G}) applies in the
majority of cases and $H(q)= \frac{\sin(q\gamma)}{q\gamma}-1$.
When the growth rate $\lambda^\textrm{HK}_{q}$ is positive, the
homogeneous state is unstable and the subsequent evolution
gives rise to cluster formation, a situation identified with
order, whereas a negative growth rate implies that the
homogeneous state is stable and clusters can not form, a sort
of disordered state.

\subsection{Unlimited random jumps}
\begin{figure}[ht]
\begin{center}
\mbox{\includegraphics[clip,angle=270,width=.2 \textwidth]{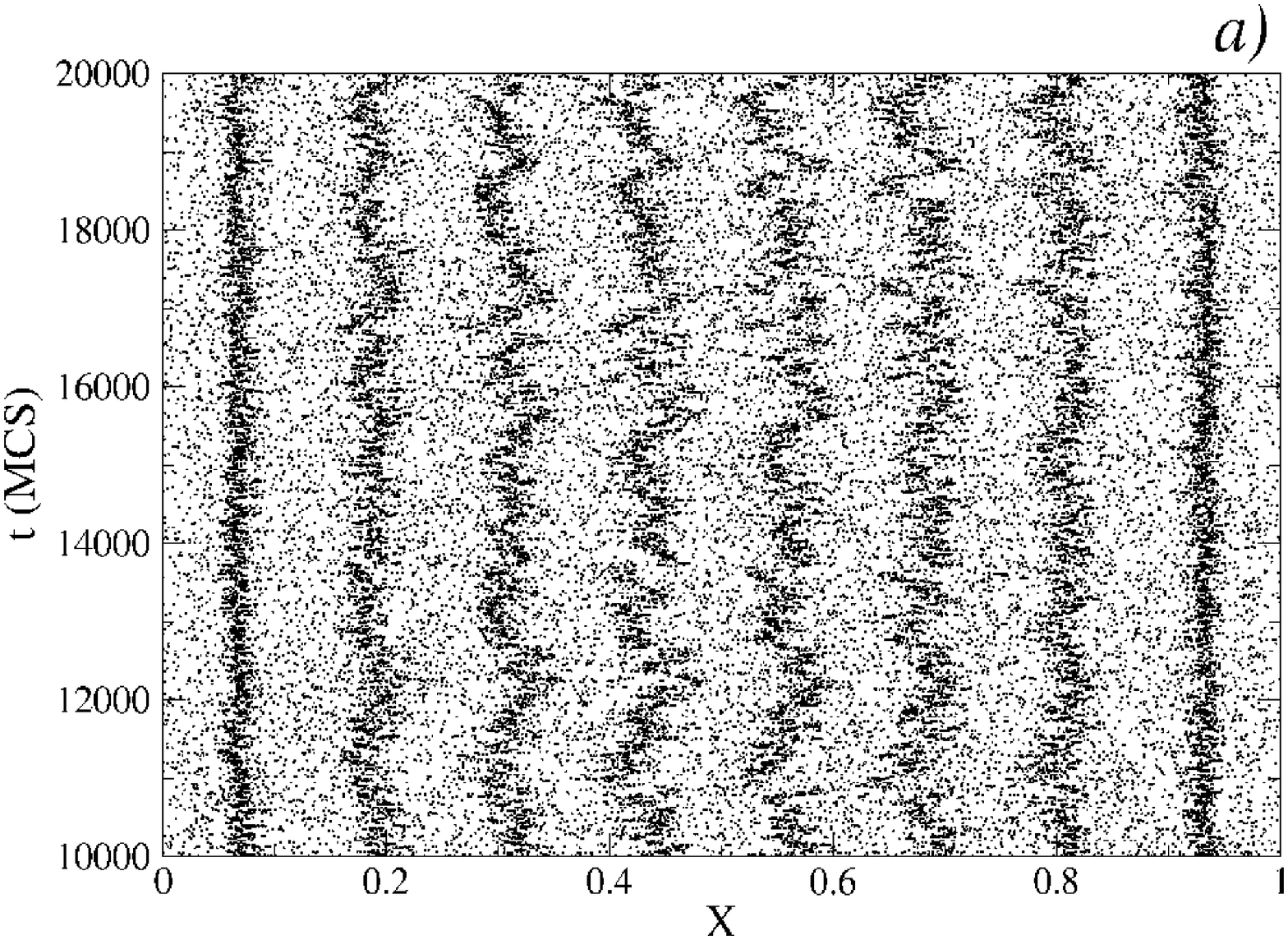}
\includegraphics[clip,angle=270,width=.2 \textwidth]{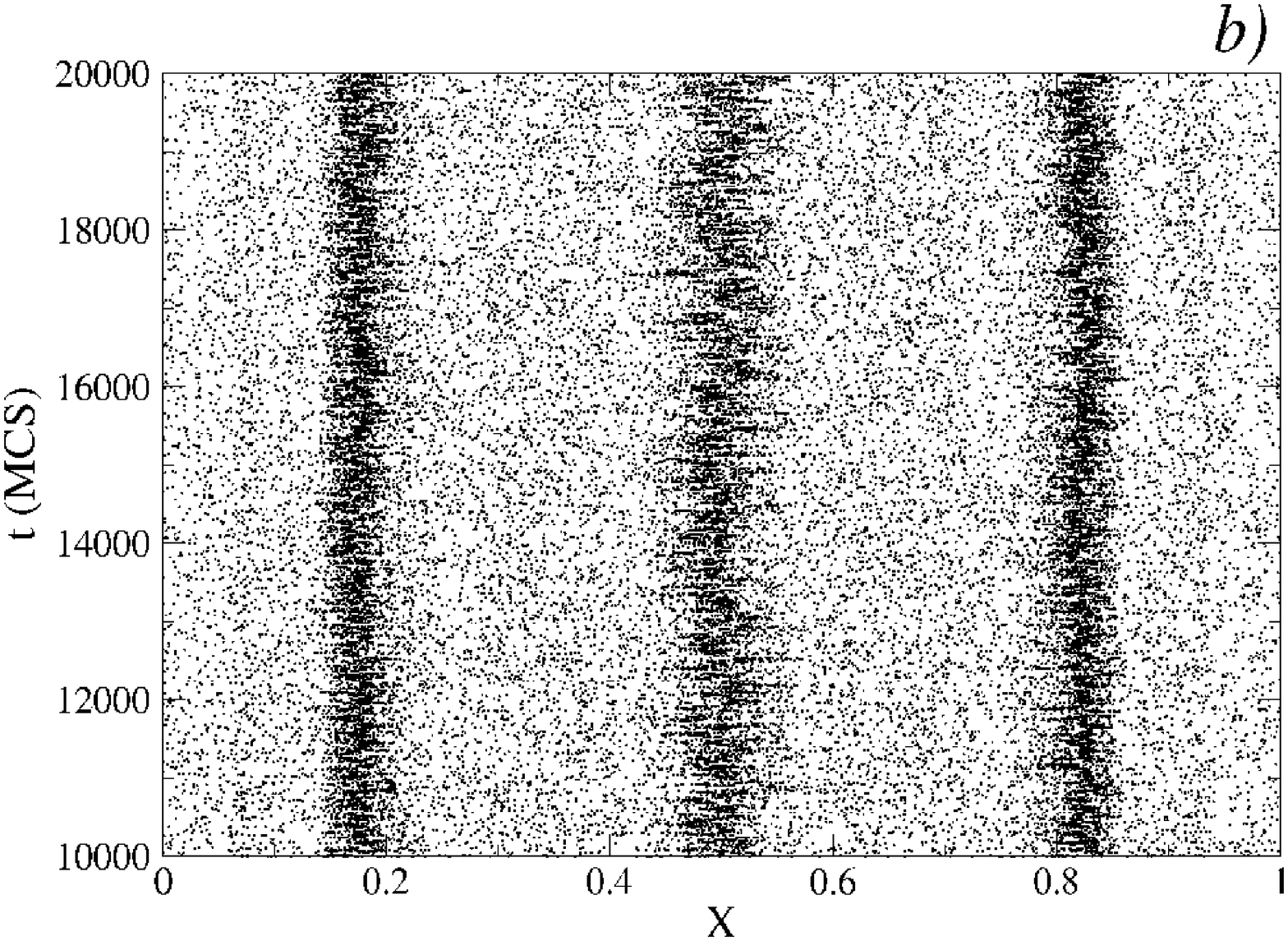}}
\hfill
\mbox{\includegraphics[clip,angle=270,width=.2 \textwidth]{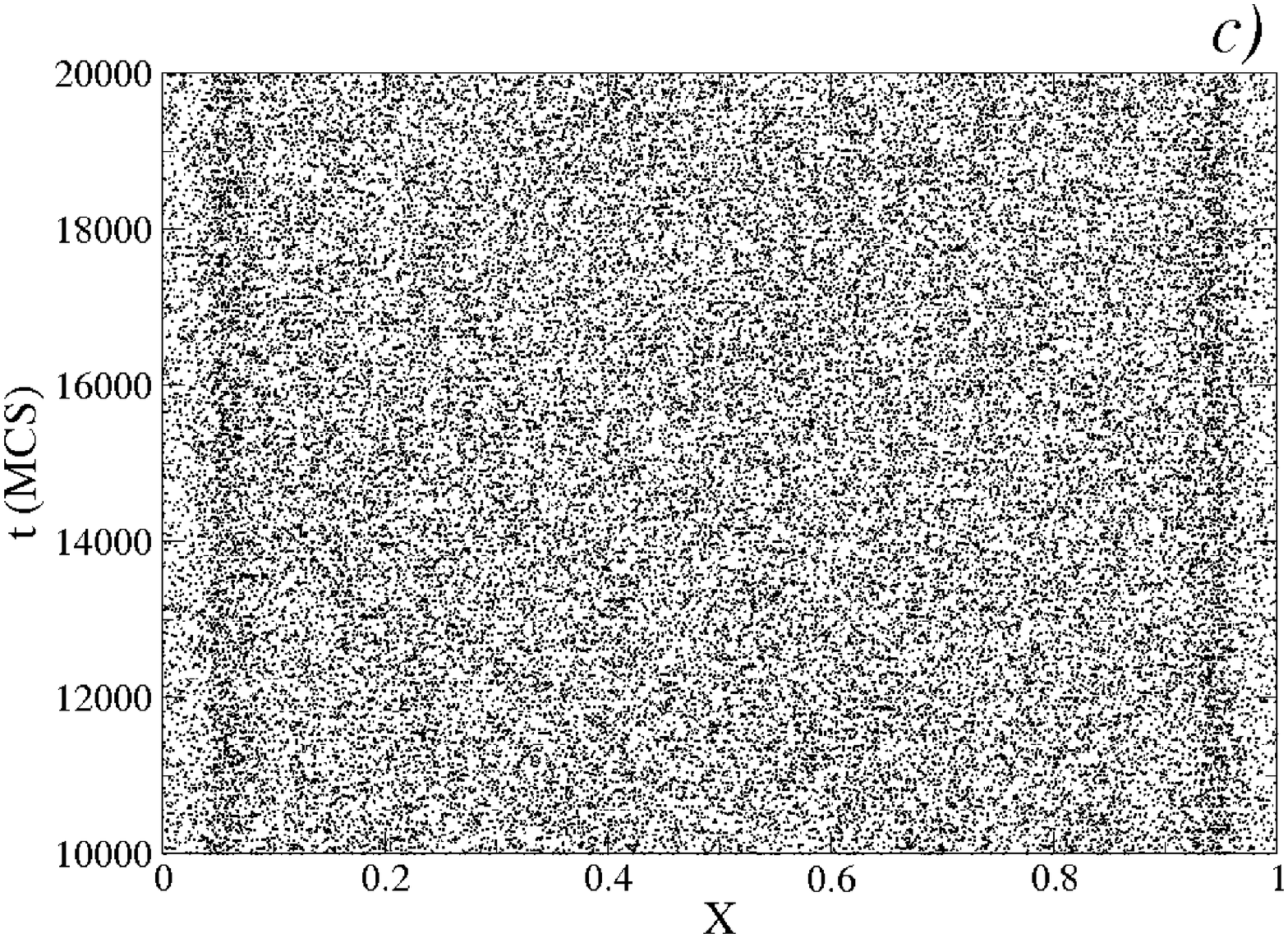}
\includegraphics[clip,angle=270,width=.2 \textwidth]{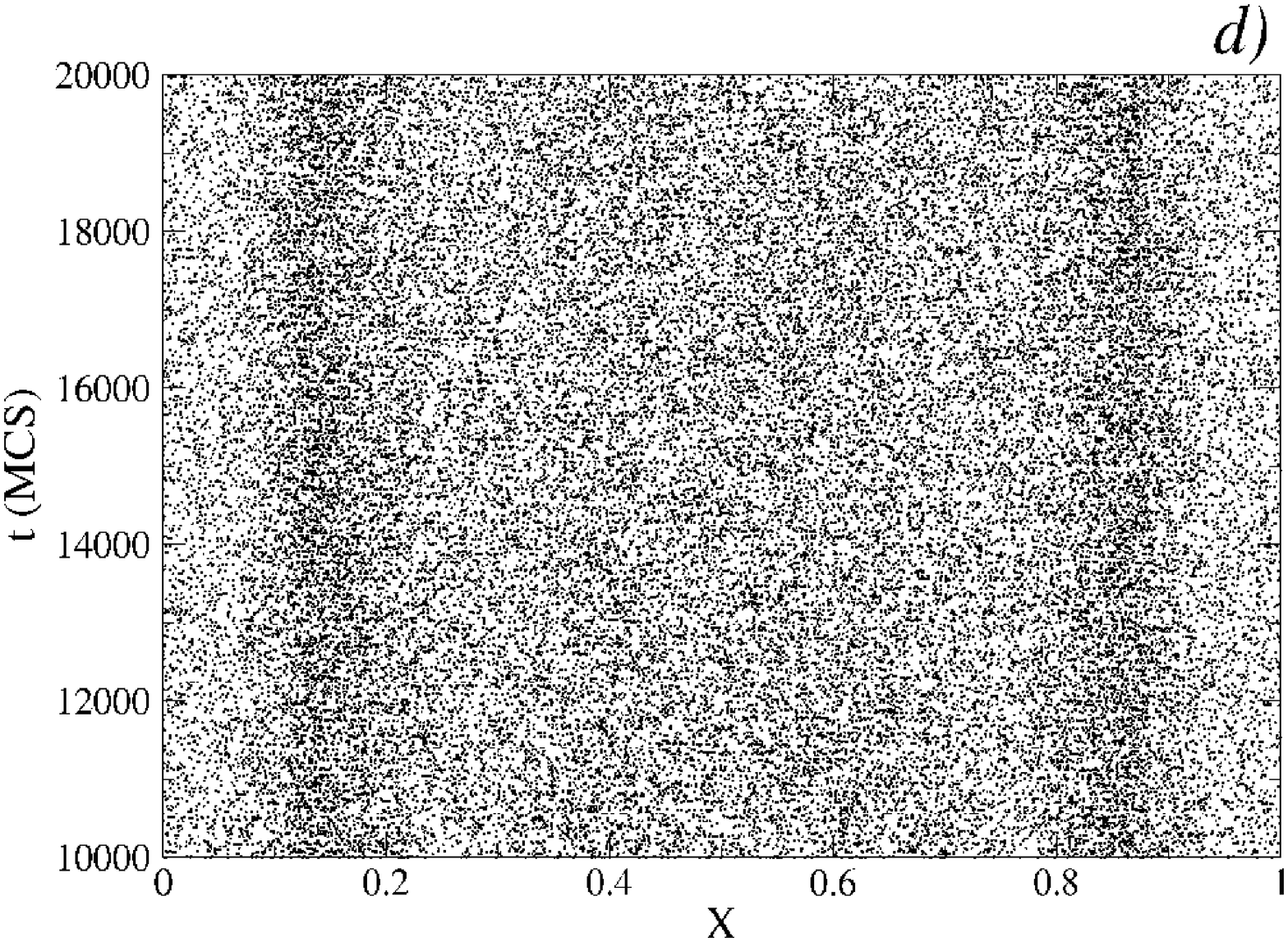}}
\end{center}
\caption{\label{fig6} Group dynamics for unlimited jumps as a function of
noise intensity $m$ and confidence parameter $\epsilon$. This figure presents
cases for $m=0.3$ (a-b), and $m=0.6$ (c-d) at $\epsilon=0.05$
(Left panels) and $0.125$ (Right panels). Its shows that for
$m>m_{c}\approx 0.51$ an unstructured state dominates (except close to the borders, where
boundary effects prevail) and opinion
clusters do  not develop. But, for $m<m_{c}$, opinion clusters
exist even for very small values of $\epsilon$. The opinion space runs from
$0$ to $L=1$ and only $100$ opinions are plotted out of $N=1000$.
Data is plotted after a long enough simulation time.}
\end{figure}

In this case, the opinion jumps are homogeneous around the
whole opinion space and therefore $H(q)=-1$. To analyze the
impact of noise on the growth rate, Fig.~\ref{fig5} shows
$\lambda^\textrm{HK}_{q}$ versus $q\epsilon$ for several values
of noise intensity $m$. From this figure, one can observe that
there is a single wavelength $q$ with the largest growth rate.
For large $m$, the maximum growth rate becomes negative, the
homogeneous state is stable and clusters do not develop. This
happens for $m>m_{c} \approx 0.51$, independently of
$\epsilon$. This result tells us that well-developed patterns
of opinion clusters are possible only for $m<m_{c}$, and that
the unstructured state is unstable in this region. The
wavenumber corresponding to the growth rate that dominates and
sets the wavelength is $q_{max}\approx2.8/\epsilon$. This gives
us an estimation of the number of opinion clusters by
recognizing that the associated periodicity is $2\pi/q_{max}$
and then the number of clusters in the unit interval is
$n_\textrm{HK} \approx 0.4/\epsilon$ ($n_\textrm{HK} \approx
0.4 L/\epsilon$ in the $[0,L]$ interval). These conclusions can
be verified in Fig.~\ref{fig6} which shows time series from
Monte Carlo simulations. For strong noise, $m>m_{c}$,
perturbations decay with time and the uniform state is
restored. Whereas, for weak noise intensity, $m<m_{c}$,
perturbations are magnified and patterns of opinions are
established. This result also means that opinion clusters would
be still observed for very small values of $\epsilon$, if $m
<m_{c}$.

\subsection{Bounded random jumps}
\begin{figure}[ht]
\begin{center}
\mbox{\includegraphics[clip,angle=270,width=.2 \textwidth]{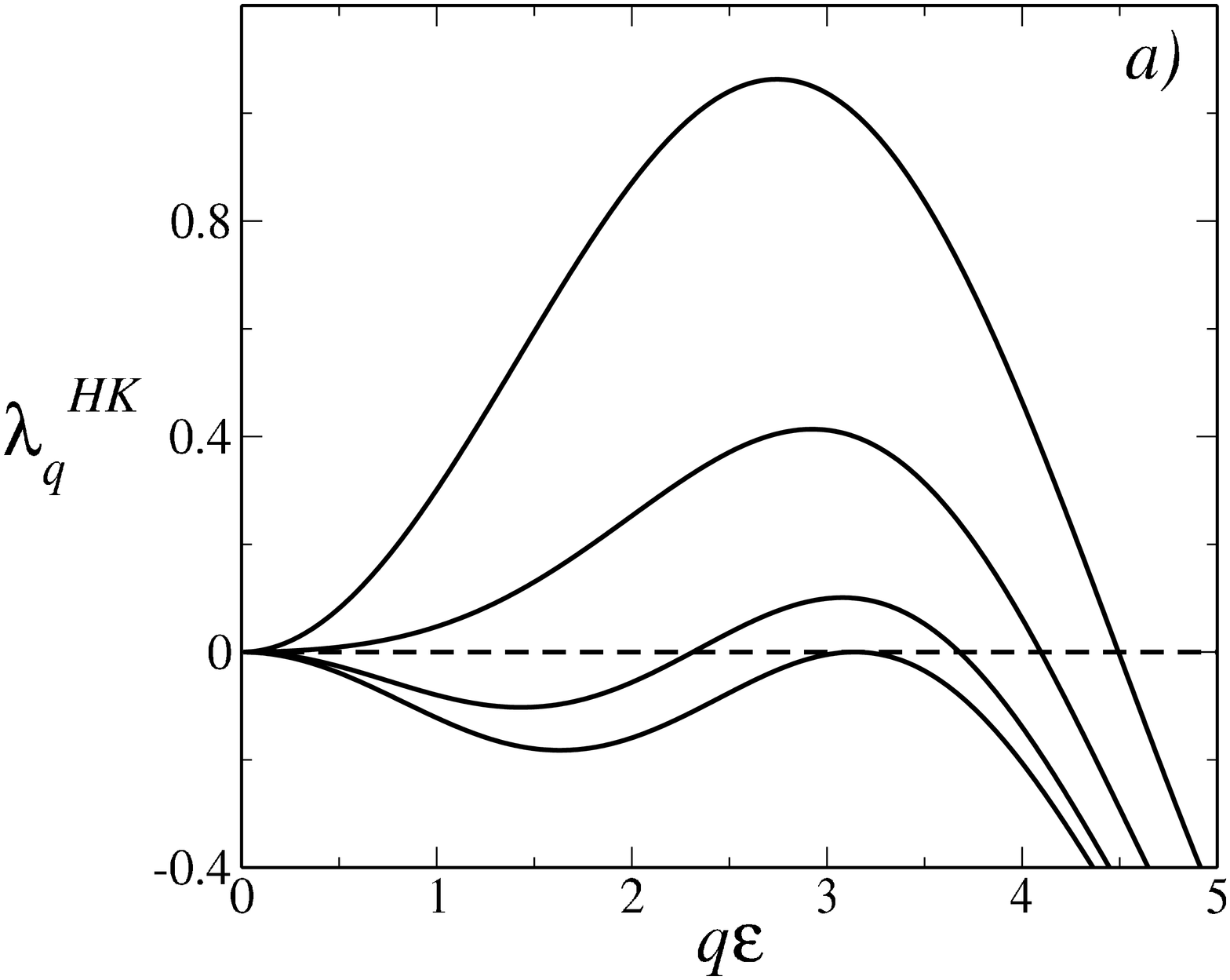}
\includegraphics[clip,angle=270,width=.2 \textwidth]{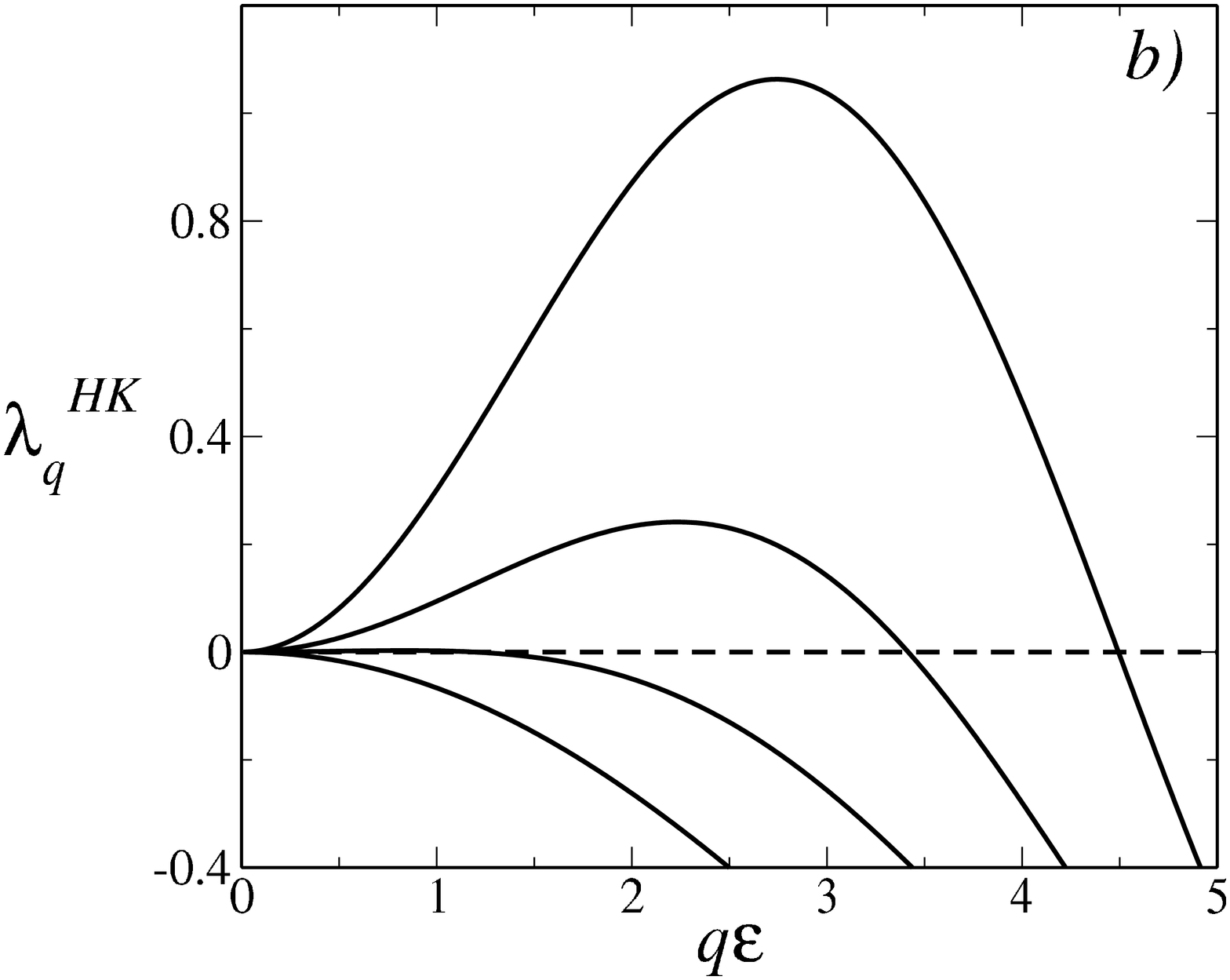}}
\hfill
\mbox{\includegraphics[clip,angle=270,width=.2 \textwidth]{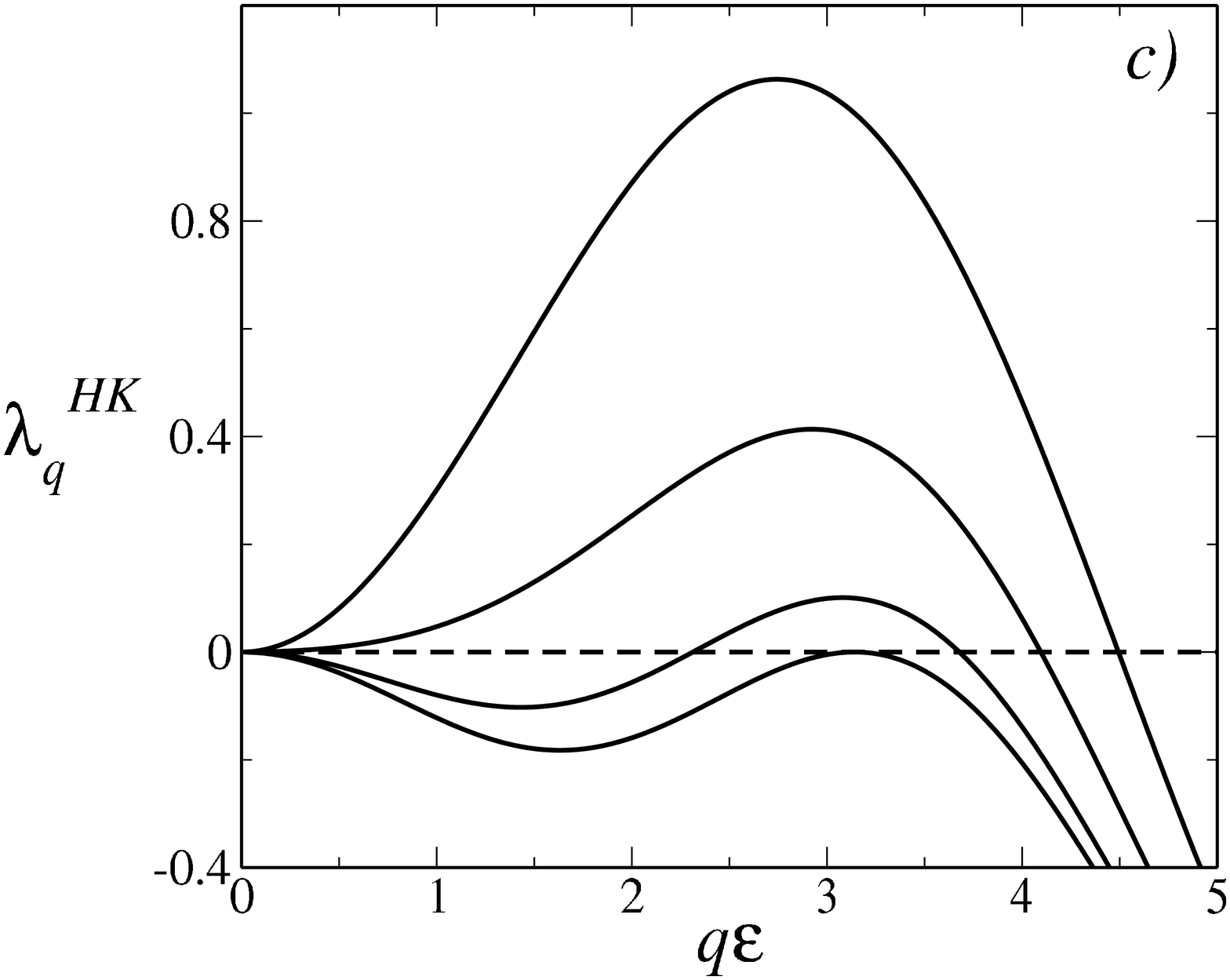}
\includegraphics[clip,angle=270,width=.2 \textwidth]{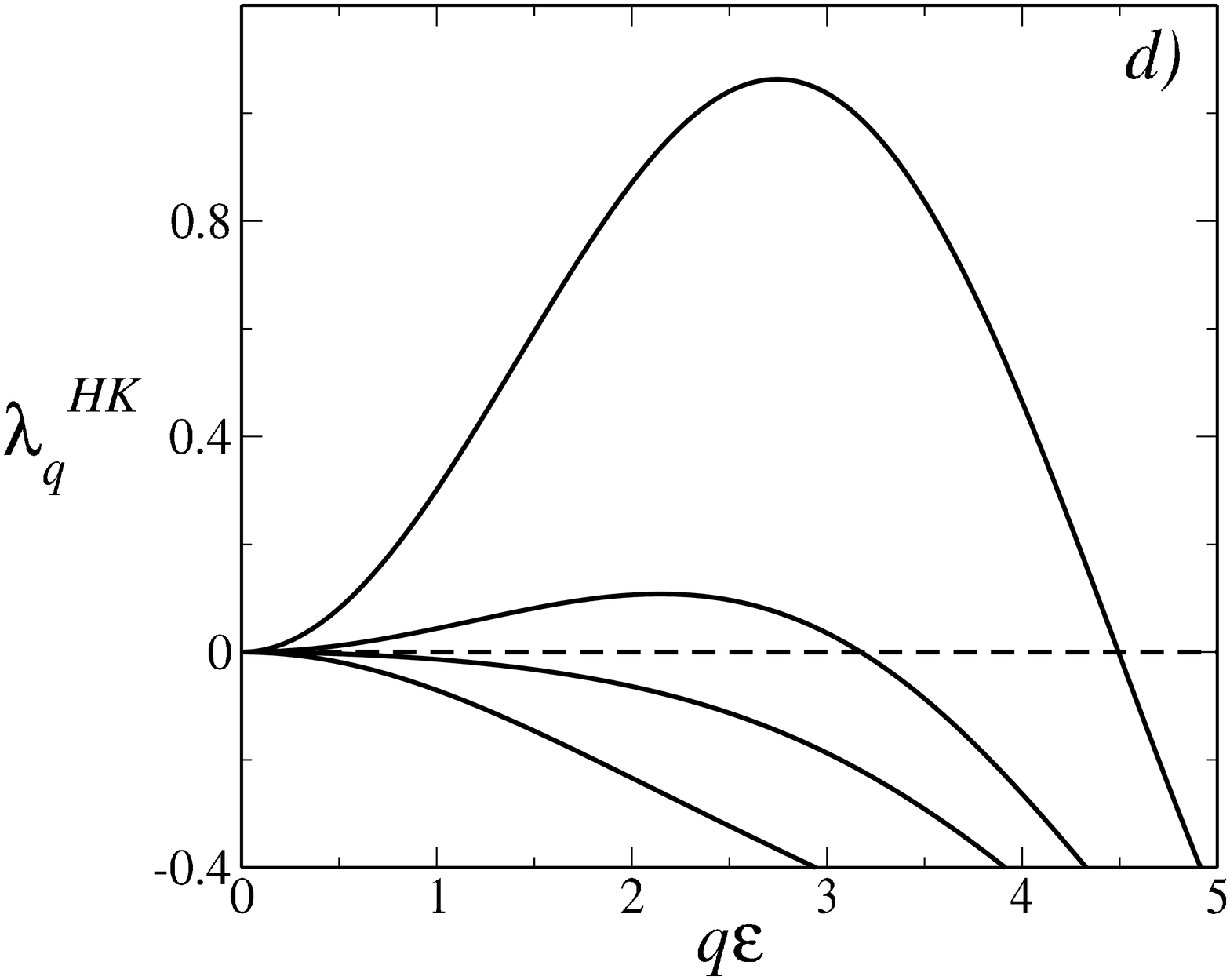}}
\end{center}
\caption{\label{fig7} Growth rate for the case of bounded random jumps.
Panels (a and b) show $\gamma=0.1$. In a) the growth rate for cases
$\epsilon<\epsilon_{c} \approx 0.068$ is presented with $m=0.0$, $0.3$, $0.45$, $0.5$,
from top to bottom. Growth rate becomes positive at a well-defined non-zero $q$.
In b) the growth rate for cases $\epsilon>\epsilon_{c} \approx 0.068$ is presented
with $m=0.0$, $0.45$, $0.65$, $0.8$, from top to bottom. In this situation, the
appearance of positive values occurs first at values of $q$ close to zero and
therefore an expansion in powers of $q$ is possible. Panels (c and d) show the
same but for $\gamma=0.4$, where $\epsilon_{c} \approx 0.28$. In c) $m=0.0$, $0.3$, $0.45$, $0.5$,
from top to bottom. In d) $m=0.0$, $0.45$, $0.55$, $0.65$, from top to bottom.}
\end{figure}
In this case the confidence mechanism is generalized by
allowing individuals to change their opinions randomly inside a
small interval $[-\gamma,\gamma]$ centered at the current
opinion. As mentioned before, the growth rate in this case
involves $H(q)= \displaystyle\frac{\sin(q\gamma)}{q\gamma}-1$
which can also be written as $H(q)=
\displaystyle\frac{\sin(q\epsilon\gamma/\epsilon)}{q\epsilon\gamma/\epsilon}-1$
to stress the dependence on $\epsilon$ when the growth rate is
plotted as a function of $q\epsilon$. Figure~\ref{fig7} shows
that the growth rate for a given $\gamma$ exhibits two regimes
as a function of $\epsilon$. For $\gamma=0.1$ and $0.4$, the
critical transitions between these two regimes are located at
$\epsilon_{c} \approx 0.068$ and $\epsilon_{c} \approx0.28$,
respectively. Figure~\ref{fig7}(a) shows the shape of
$\lambda^\textrm{HK}_{q}$ as a function of $q\epsilon$ for
$\gamma=0.1$ and $\epsilon<\epsilon_{c}\approx 0.068$. The form
of this growth rate allows us to conclude that the perturbation
with the largest growth rate dominates. However, the maxima of
$\lambda^\textrm{HK}_{q}$ and the appearance of positive values
must be obtained numerically. On the other hand, for
$\epsilon>\epsilon_{c}$ [see Fig.~\ref{fig7}(b)], the
appearance of positive values of $\lambda^\textrm{HK}_{q}$ when
varying the noise intensity occurs first at values of $q$ close
to zero corresponding, as expected for these large values of
$\epsilon$, to a long-wavelength instability. In this limit,
approximate analytical expressions can be obtained expanding
$\lambda^\textrm{HK}_{q}$ in powers of $q$:
\begin{equation}
\lambda^\textrm{HK}_{q}=\frac{(1-m)\left(1-\mu\right)\epsilon^2}{3}q^{2}-
\frac{4(1-m)\epsilon^4}{5!}q^4 +{\cal O}(q^6),
\label{eq:expansion}
\end{equation}
where $\mu=\frac{m\gamma^2}{(1-m)\epsilon^2}$. Because the
$q^4$ term is always negative, the change of the sign of the
$q^2$ term identifies
\begin{equation}
m_{c}=\frac{2\epsilon^2}{2\epsilon^2+\gamma^2}
\label{eq:mc}
\end{equation}
as the value below which opinion clusters appear. Within this
approximation and close to the instability threshold the
fastest growing mode is:
\begin{equation}
q_{max} \approx \frac{\sqrt{5}}{\epsilon}\left(1-\mu\right)^{1/2}.
\label{eq:mode}
\end{equation}
\begin{figure}[ht]
\begin{center}
\mbox{\includegraphics[clip,angle=270,width=.3 \textwidth]{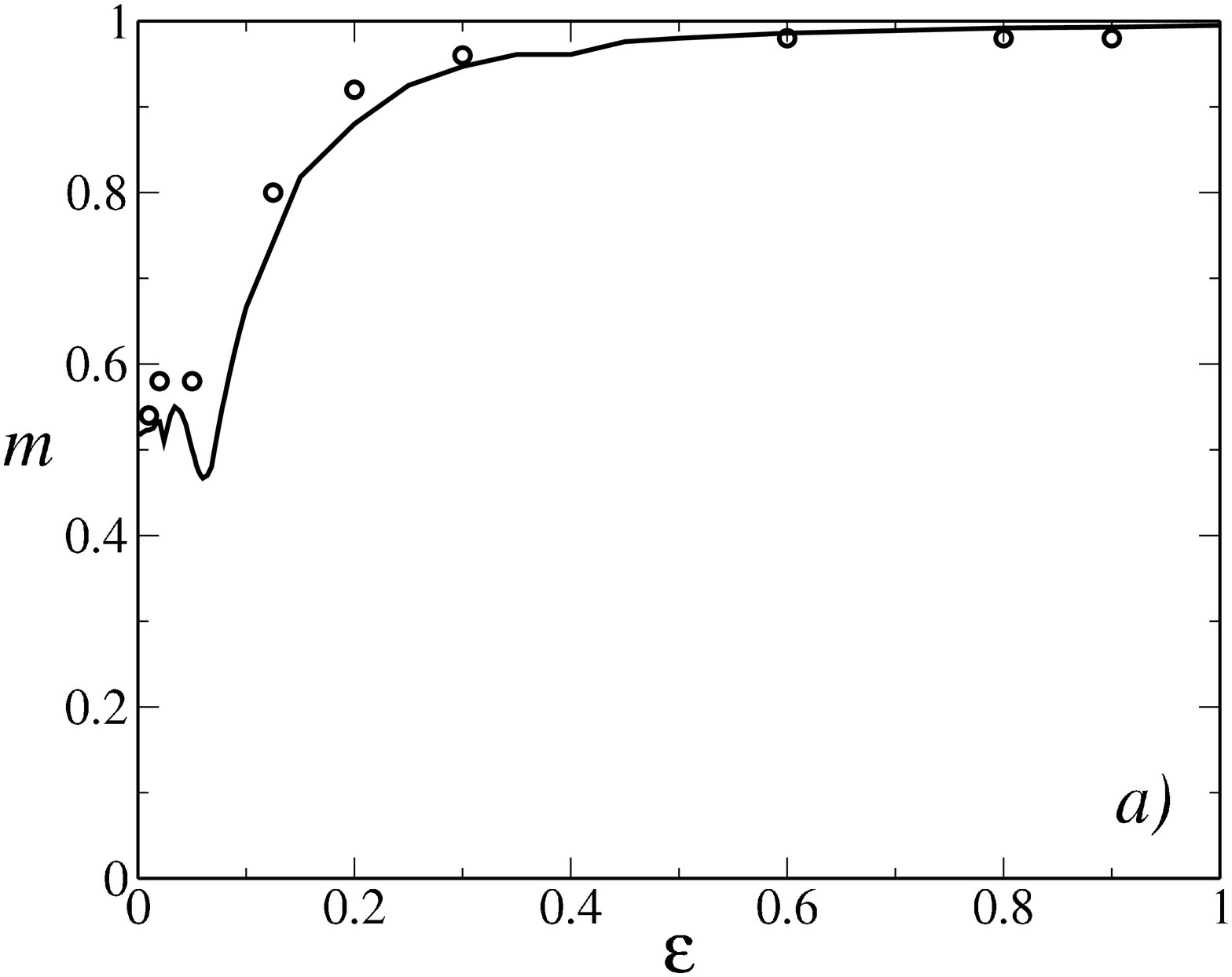}}
\end{center}
\begin{center}
\mbox{\includegraphics[clip,angle=270,width=.3 \textwidth]{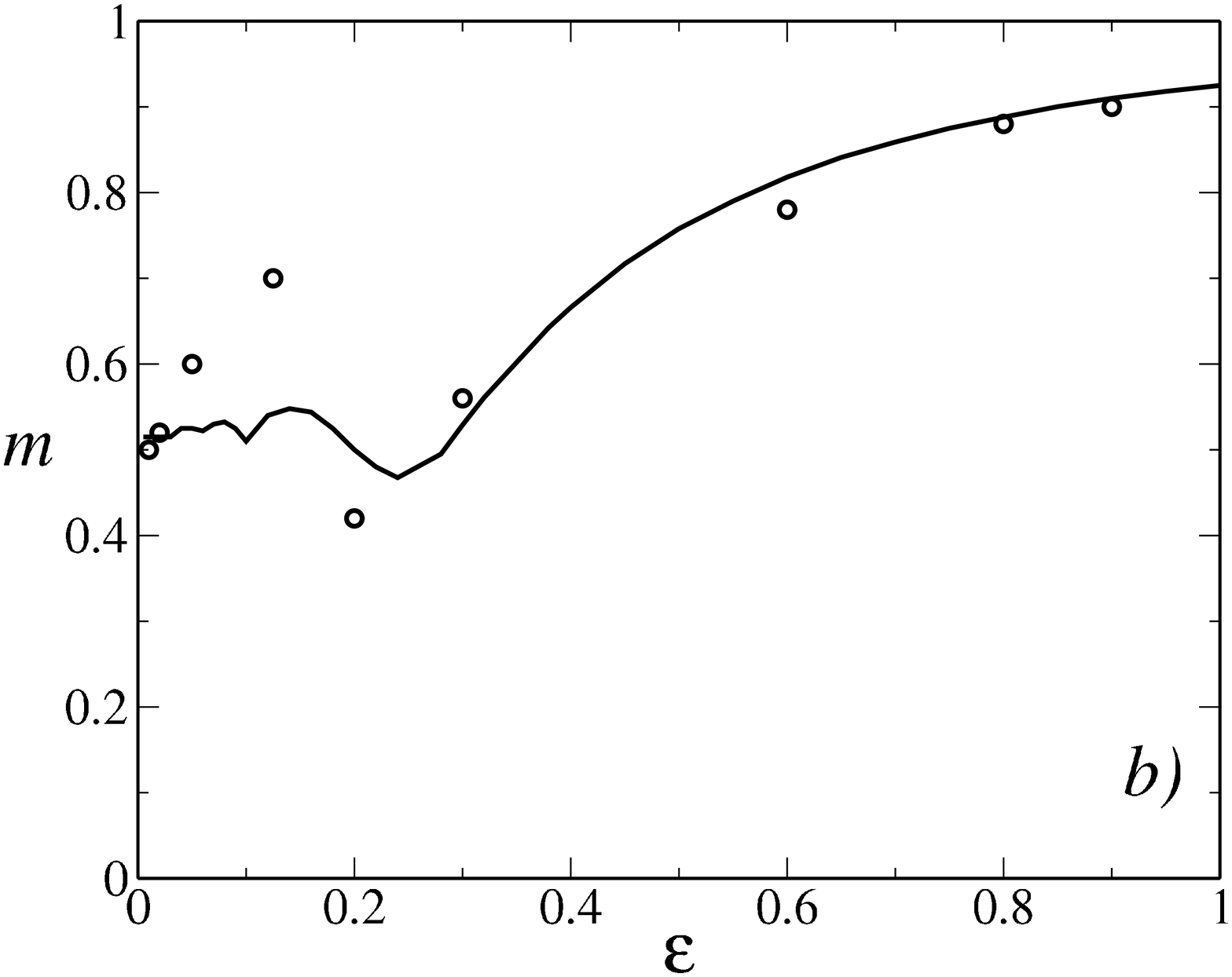}}
\end{center}
\caption{\label{fig8} Phase diagram on the plane ($m,\epsilon$) for the case
of bounded jumps obtained from our linear stability analysis (solid lines) and
compared with the results coming from the occurrence of
the maximum value of the cluster coefficient $G_M$ as a function of $m$ for fixed $\gamma$,
obtained from Monte Carlo simulations using adsorbing boundary conditions
and $N=10^{4}$ (open dots). Clusters appear below these lines, whereas the disordered state is stable
above. (a) $\gamma=0.1$. (b) $\gamma=0.4$. For $\epsilon<\epsilon_c=0.068$ (case $\gamma=0.1$) and
$\epsilon<\epsilon_c=0.28$ (case $\gamma=0.4$), the solid line is obtained numerically
from the change of sign of the maximum of the growth rate, Eq. (\ref{eq:drgeneral}), but
for $\epsilon>\epsilon_{c}$ the approximate expression (\ref{eq:mc}) is used, which is virtually identical.
In this phase diagram $L=1$. }
\end{figure}
Figures ~\ref{fig7}(c) and (d) show that the situation is
similar for $\gamma=0.4$. Figure~\ref{fig8} presents the
critical lines for existence of opinion clusters in the
parameter space $(m,\epsilon)$ for the cases considered in this
work. In this case, to identify in a more quantitative way the
order-disorder transition from Monte Carlo simulations with
adsorbing boundary conditions, we use the so-called cluster
coefficient $G_{M}$ \cite{pineda1,pineda2}. One divides space
$[0,1]$ in $M$ equal boxes and counts the number of individuals
$l_i$ which, at time step $n$, have their opinion in the box
$[(i-1)/M,i/M]$. We choose $M=100$. Then, one defines an
entropy $S_M=-\sum_{i=1}^{M}\frac{l_i}{N}\ln\frac{l_i}{N}$,
from which the cluster coefficient is defined as
\begin{equation}
G_M=M^{-1}\left\langle e^{\overline S_M}\right\rangle,
\label{eq:CC}
\end{equation}
where the over-bar denotes a temporal average in the long-time
asymptotic state and $\langle \cdot\rangle$ indicates an
average over different realizations of the dynamics. Note that
$1/M\le G_M\le 1$. Large values of $G_M$ indicate a situation
identified with disorder, while small values of $G_M$ indicate
that opinions peak around a finite set of major opinion
clusters (a situation identified with order). The adsorption by
the borders prevents the fully homogeneous state $G_{M}=1$, as
two opinion clusters are always formed at the extremes.
Therefore, we will consider that the transition from order to
disorder is the location $m_{c}$ of the maximum value of
$G_{M}$ for fixed $\epsilon$ and $\gamma$ (results plotted in
Fig. \ref{fig8}).

\section{Comparison with the noisy DW model}
\label{sec:4}
The bounded confidence mechanism by which two individuals only
influence the opinion of each other if their respective
opinions differ less than some given amount holds for the DW
model and the HK model. The HK model only differs from the DW
model in that the interactions take place in groups rather than
in pairs. In the noiseless DW model, one starts with a random
distribution in opinion space $[0,L]$ and at subsequent time
steps two randomly chosen agents may change their opinions to
the average of both opinions if their opinions differ less than
some given amount $\epsilon$ (in the standard particular case
in which a convergence parameter in the model is equal $0.5$).
A detailed analysis of this model shows that the bifurcation of
opinion clusters as a function of $\epsilon$ differs
quantitatively from the noiseless HK model \cite{lorenz1}. For
instance, they have different critical values of $\epsilon$ for
the consensus transition and, unlike the HK model, the DW model
exhibits low-populated opinion clusters at the extremes and
between major clusters even for uniform initial conditions.
Nevertheless, they are similar in the fact that the bifurcation
and nucleation of clusters observed outside the consensus region 
seems to repeat itself for decreasing $\epsilon$ in such a way
that intercluster distances scale approximately with
$1/\epsilon$. On the other hand, the DW model has been also
studied under opinion random jumps and interesting phenomena
arising from this randomness have been reported
\cite{pineda1,pineda2,jensen1}. This section will be devoted to
compare the results presented in previous sections with those
observed in the noisy DW model.

The first conclusion we arrive is that, under unlimited opinion
jumps, both models exhibit low-populated opinion clusters at
the extremes and between high populated clusters. We also
observed in both cases that the number of individuals belonging
to these clusters increases when the noise intensity increases.
The remarkable fact of bistability regions, reported first in
the noisy DW model \cite{pineda1}, is also observed in the
noisy HK model. Inside these regions one finds that inherent
fluctuations arising from the finite number of individuals take
the system from one state to the other and back.

The coarsening process, observed in the noisy DW model
\cite{pineda2}, is also presented in the HK model when bounded
random jumps of opinions are allowed to occur just inside a
very small interval centered at the moment opinion. Clusters
seem to perform a kind of random walk in opinion space and they
merge when they collide. When the interval where jumps occur is
larger, we observed that like the DW model the HK model admits
a stable pattern of opinion clusters with reduced wandering and
with regions of bistability. But it seems that in the HK model
it is harder to find  multiple jumps between one state to
another and back. We just found jumps from one state to another
but the new state never comes back to the previous one.
\begin{figure}[ht]
\begin{center}
\mbox{\includegraphics[clip,angle=270,width=.3 \textwidth]{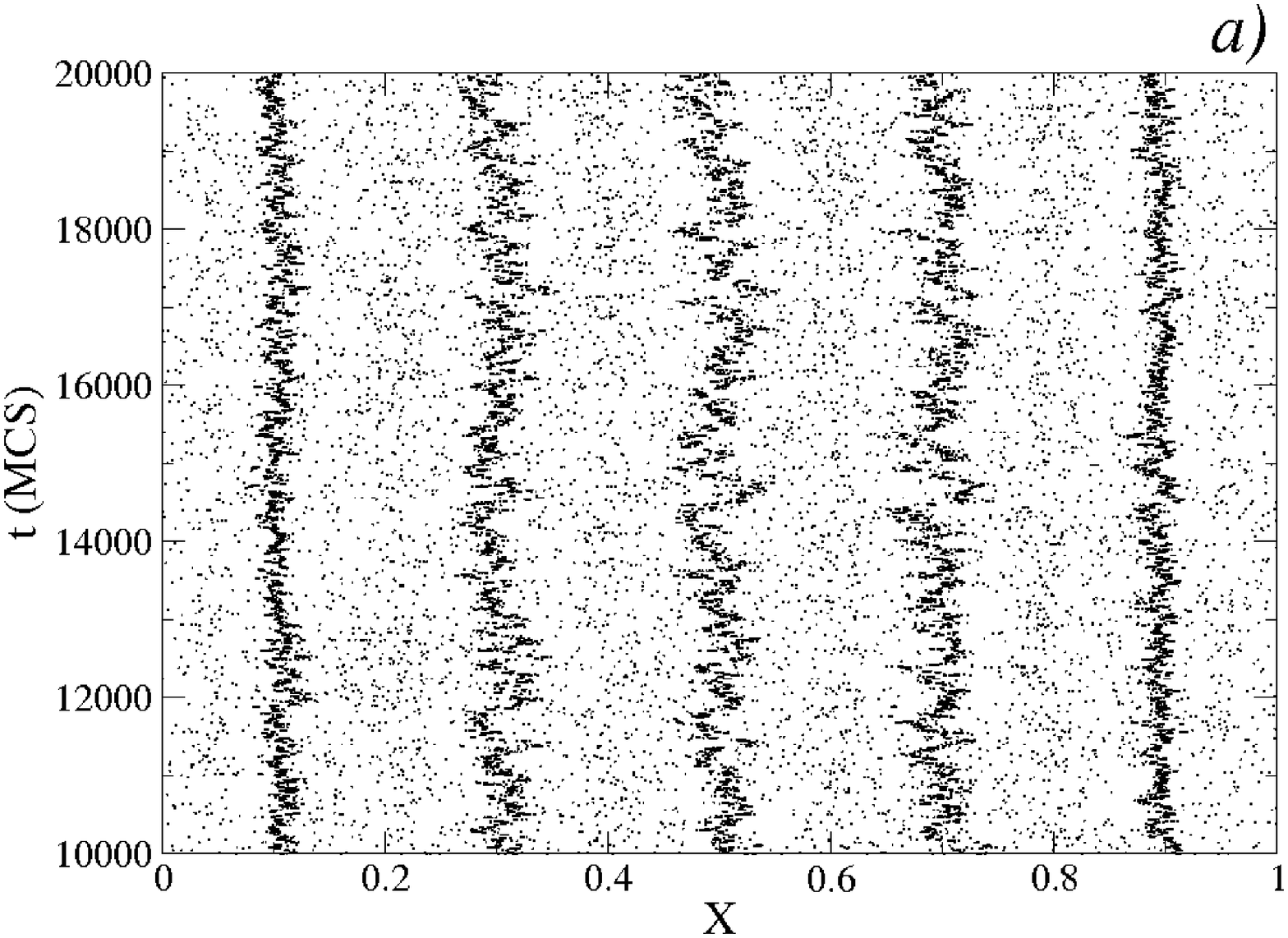}}
\end{center}
\begin{center}
\mbox{\includegraphics[clip,angle=270,width=.3 \textwidth]{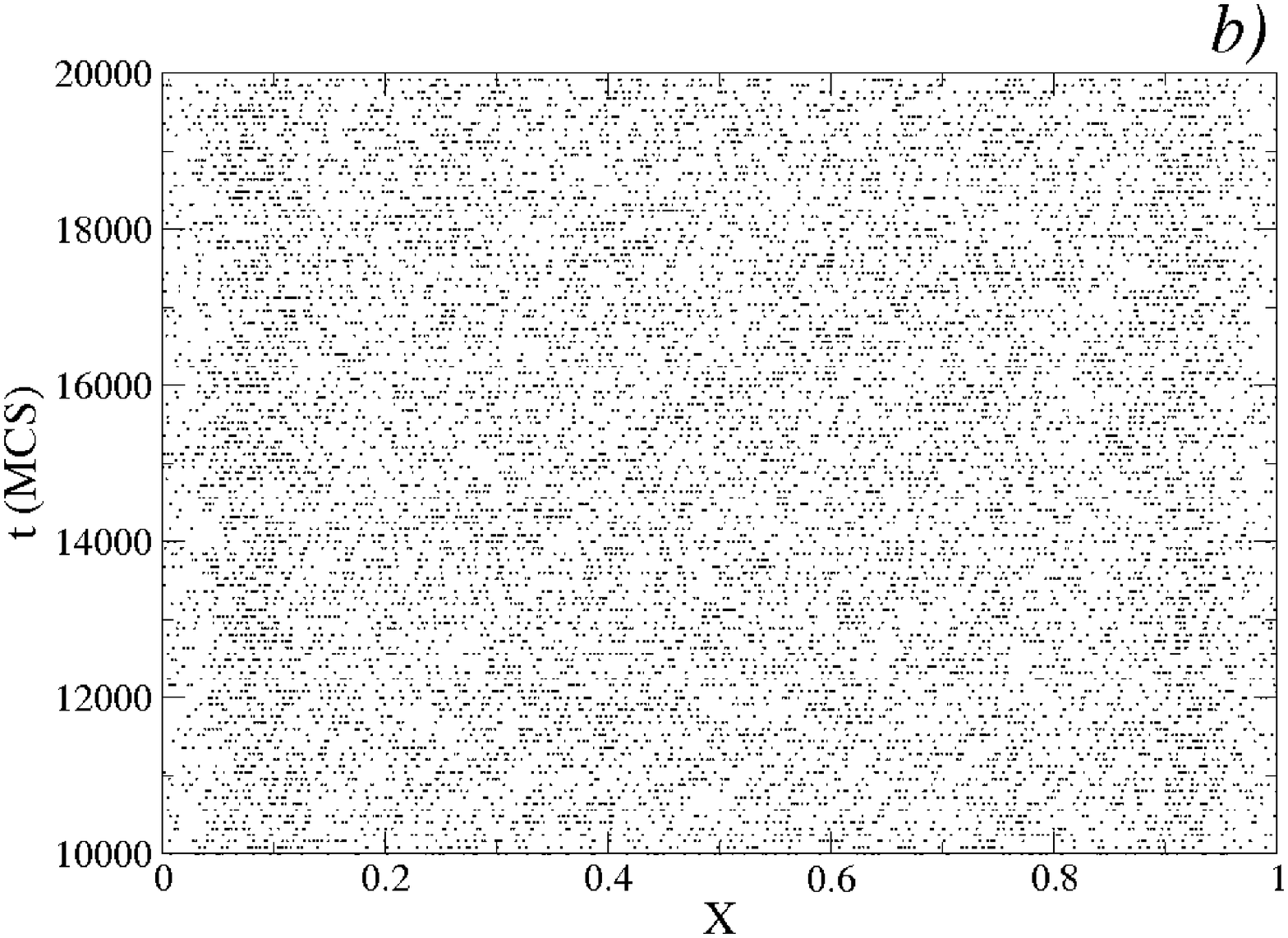}}
\end{center}
\caption{\label{fig9} Opinion dynamics of the HK model (a) and DW model (b) for
noise intensity $m=0.1$ and confidence parameter $\epsilon=0.08$ in the
case of unlimited jumps. In this case $L=1$ and only $100$ opinions are plotted
out of $N=1000$. For the DW model, $\epsilon_{c}(m=0.1,L=1) \approx 0.096$.
Our analytical calculations predict that for the HK model the formation of patterns of opinion clusters
occurs even for these small values of $\epsilon$, but that this does not occur for DW, as actually seen
in the plots. Data starts to be plotted after long enough simulation time}
\end{figure}

The order-disorder transition as a function of noise intensity
$m$ is also observed in both models \cite{pineda1,pineda2}.
Nevertheless, the linear stability analysis revealed some
important differences that we would like to discuss in the rest
of this section. The linear stability analysis of the
unstructured solution of the DW's density-based master equation
under both types of noises and with the opinion space being
$[0,1]$ with periodic boundary conditions gives for the growth
rate
\begin{equation}
\lambda^\textrm{DW}_{q}=4(1-m)\epsilon\left[
\frac{4\sin(q\epsilon/2)}{q\epsilon}-\frac{\sin(q\epsilon)}{q\epsilon}-1
\right] + m H(q).
\label{eq:q8}
\end{equation}
$H(q)$ is, for both types of noise, the same function as in the
HK case. This result clearly shows that, unlike the HK model, the
first term of the growth rate $\lambda^\textrm{DW}_{q}$ carries
as a prefactor the confidence parameter $\epsilon$. This
difference makes the time scales between the two models to be
different, and slows down the DW instability for small $\epsilon$.
Since the result for a different value of $L$ is recovered by
replacing $\epsilon$ by $\epsilon/L$, we also conclude that
unlike the HK model a faster instability is expected for the DW
model in smaller opinion spaces. More importantly, since the
order-disorder transition is determined by a balance between
the $m$ and the $1-m$ terms in Eq. (\ref{eq:q8}), the critical
noise value $m_c$ below which there is opinion cluster
formation is now a function of $\epsilon$ for both type of noise,
at variance with the HK case.
\begin{figure}[ht]
\begin{center}
\mbox{\includegraphics[clip,angle=270,width=.22 \textwidth]{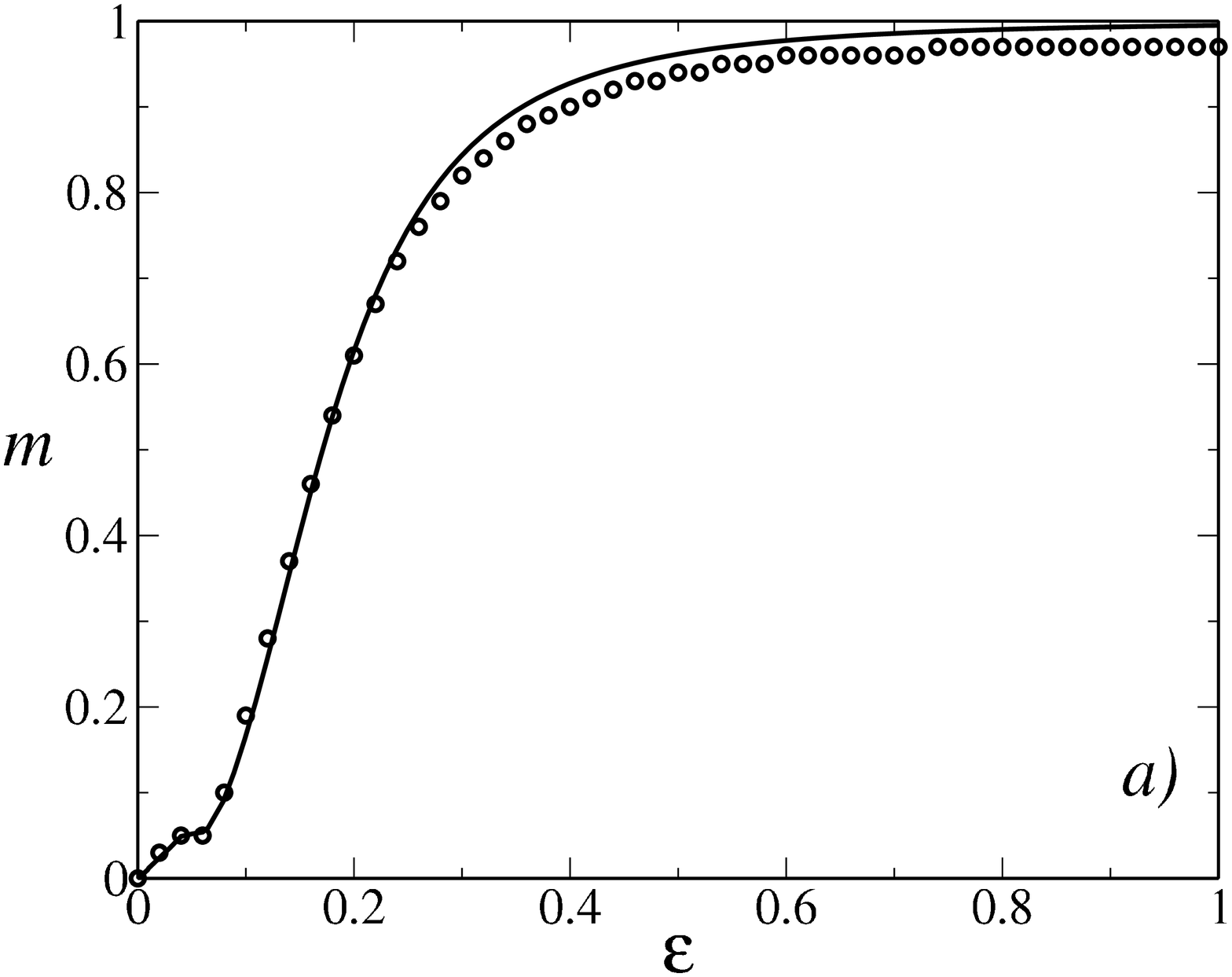}}
\end{center}
\begin{center}
\mbox{\includegraphics[clip,angle=270,width=.22 \textwidth]{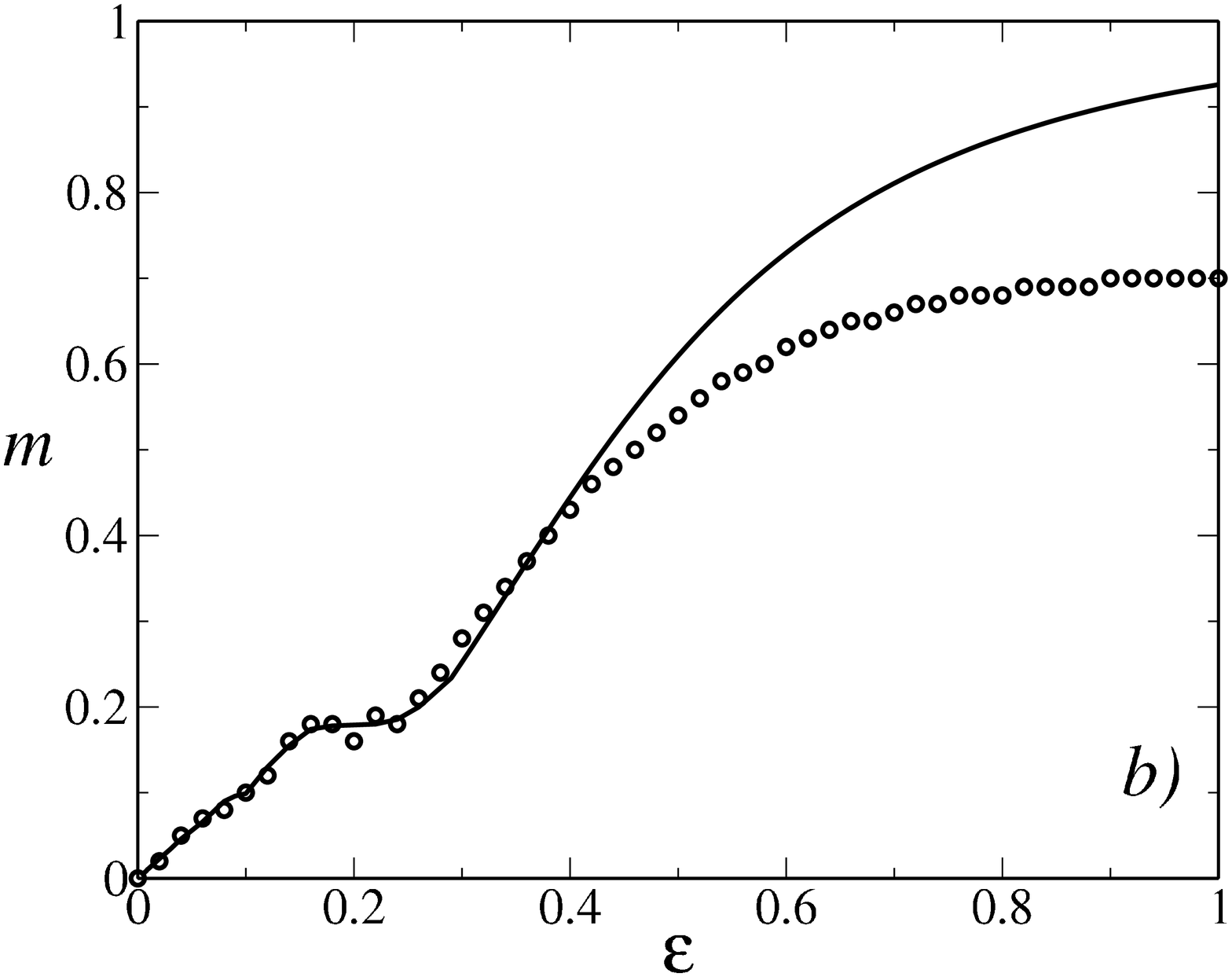}}
\end{center}
\caption{\label{fig10} Phase diagram of the DW model on the plane ($m,\epsilon$) for the case
of bounded jumps obtained from our linear stability analysis (solid lines) and
compared with the results coming from the occurrence of
the maximum value of the cluster coefficient $G_M$ as a function of $m$ for fixed $\gamma$,
obtained from Monte Carlo simulations using adsorbing boundary conditions
and $N=10^{4}$ (open dots). Clusters appear below these lines, whereas the disordered state is stable
above. (a) $\gamma=0.1$. (b) $\gamma=0.4$. For $\epsilon<\epsilon_c=0.076$ (case $\gamma=0.1$) and
$\epsilon<\epsilon_c=0.31$ (case $\gamma=0.4$) the solid line is obtained numerically
from the change of sign of the maximum of the growth rate, Eq. (\ref{eq:q8}), but
for $\epsilon>\epsilon_{c}$ the approximate expression (\ref{eq:mcDW}) is used, which is virtually identical.
In this phase diagram $L=1$. This figure should be compared with Fig. 7 in \cite{pineda2} and Fig.~\ref{fig8} of this work. 
The discrepancies between lines and dots at large $\epsilon$ arise from the influence of the adsorbing boundary conditions of the
Monte Carlo case, whereas the analytical calculations assume periodic boundary conditions.}
\end{figure}

For the case of unlimited random jumps $[H(q)=-1]$ one finds
that the maximum value of $\lambda^\textrm{DW}_{q}$ is negative
for $m>m_{c}$ and positive for $m<m_{c}$, where
$m_{c}\approx\epsilon/(0.8676+\epsilon)$. Alternatively, for
fixed $m$ the maximum growth rate is negative for
$\epsilon<\epsilon_{c}$ and positive for
$\epsilon>\epsilon_{c}$, where
$\epsilon_{c}\approx0.8676m/(1-m)$ \cite{pineda1}. The absolute
maximum of the growth rate occurs at
$q_{max}\approx2.8/\epsilon$, similar to the one of the HK
model. It means that the number of clusters predicted as a
function of the control parameters is, for both models,
$n_{DW}=n_{HK} \approx 0.4L/\epsilon$. Note that while in the
HK model one observes the formation of patterns of opinion
clusters for small values of $\epsilon$ if noise intensities
are small ($m <m_{c}=0.51$), in the DW model there is a minimal
value of $\epsilon_{c}(m,L)$ below which opinion clusters do
not appear. Figure~\ref{fig9} shows Monte Carlo simulations
that verify these results for $m=0.1$ and $L=1$. With these
parameter values we get that the critical condition for cluster
formation in the DW model is $\epsilon_{c}(m=0.1,L=1) \approx
0.096$. It means that for $\epsilon=0.08$ the homogeneous state
dominates. But, as predicted above, in the noisy HK model
clusters are still possible for these parameter values. In
fact, the number of opinion clusters predicted is $n_{HK}
\approx 5$, in agreement with the numerical results displayed
in this same figure.

For bounded random jumps of opinions, we observe that like in
the HK model the growth rate of the DW exhibits two regimes for
fixed $\gamma$ while $\epsilon$ varies. In fact, for
$\gamma=0.1$ and $0.4$, the critical transitions between
regimes are given by $\epsilon_{c } \approx 0.076$ and $0.31$,
respectively. Similar to the HK model, for
$\epsilon<\epsilon_{c}$ the critical line must be obtained
numerically. For $\epsilon>\epsilon_{c}$ the appearance of
positive values of $\lambda^\textrm{DW}_{q}$ occurs first at
values of $q$ close to zero, identifying again a long-wave
instability, and one can find an approximate analytical
expression for the critical condition given by
\begin{equation}
m_{c}^{DW}=\frac{2\epsilon^3}{2\epsilon^3+\gamma^2}
\label{eq:mcDW}
\end{equation}
[compare with Eq. (\ref{eq:mc})]. The phase diagram for the
order-disorder transition in the DW model in the parameter
space $(\epsilon,m)$ is shown in Fig.~\ref{fig10}, revealing
some differences with the corresponding diagram for HK model
(Fig. \ref{fig8}), especially important at small $\epsilon$.

\section{Conclusions}
\label{sec:5}
In this paper, we have analyzed the Hegselmann-Krause model for
continuous opinion dynamics under the influence of opinion
noise. More precisely, we modify the model by giving each
individual the opportunity to change, with a given probability
$m$, his opinion to a randomly selected opinion inside the
whole opinion space $[0,L]$ or inside the interval
$[\gamma,-\gamma]$, centered around the current opinion. The
final behavior, which depends of the confidence parameter
$\epsilon$, the noise intensity $m$ and the parameter $\gamma$,
is compared with the case of zero noise, and with the Deffuant
et al. model for continuous opinion dynamics under similar
types of noise.

Monte Carlo simulations have shown that, for opinion jumps
inside the whole opinion space, the noisy HK model exhibits
low-populated clusters at the extremes and between highly
populated clusters. We found that the mass of these clusters
increases as the noise intensity increases. Similar to the
noisy DW model, we also found regions of bistability where the
fluctuations present in Monte Carlo simulations are able to
induce jumps from one state to another and back. For jumps
inside the interval $[\gamma,-\gamma]$, the main dynamics of
the system depends strongly on the parameter $\gamma$. For
small values of $\gamma$, wandering of the clusters occurs and
a coarsening process develops in which opinion clusters start
to collide and merge until a single cluster remains after long
time. For large values of $\gamma$, the mobility is reduced and
the collision of clusters disappears given rise to a stable
pattern of opinion clusters with certain regions of
bistability.

A density-based master equation is introduced and the
order-disorder transition induced by noise is analyzed using a
linear stability analysis of the unstructured solution of this
equation under periodic boundary conditions. We have derived
analytical conditions for opinion pattern formation for both
types of noise. We found qualitative, and in some cases even
quantitative, agreement between the analytical results and the
numerical simulations.

We analyzed in some detail the differences and similarities
between the noisy HK model and the noisy DW model. We found
that the most striking difference appeared concerning the dependency
of the critical conditions for opinion cluster formation with
the confidence parameter $\epsilon$.

Finally our work stresses that, although the HK and DW model
are similar in nature, their bifurcation behaviors and
phenomenology as a function of the control parameters present
important differences, also in the present of noise.

\section*{Acknowledgments}

M.P gratefully acknowledge support from USB-DID through the
project S1-IN-CB-010-12. This work was supported by FEDER and
MINECO (Spain), under projects FISICOS (FIS2007-60327) and
INTENSE@COSYP (FIS2012-30634), and by Comunitat Aut\`onoma de
les Illes Balears.

%

\begin{thebibliography}{}
\bibitem{stauffer1} D. Stauffer, AIP Conf. Proc, {\bf 779}, 56 (2005).
\bibitem{castellano} C. Castellano, S. Fortunato and V. Loreto, Rev. Mod. Phys, {\bf 81}, 592 (2009).
\bibitem{lorenz1} J. Lorenz, Int. J. Mod. Phys. C {\bf 18}, 119 (2007).
\bibitem{lorenz2} J. Lorenz, Complexity {\bf 15}, 43 (2010).
\bibitem{deffuant1} G. Deffuant, D. Neu, F. Amblard, and G. Weisbuch, Adv. Compl. Syst {\bf 3}, 87 (2000).
\bibitem{krause1} U. Krause, in Proc. Commun. Diference Equations, edited by S. Elyadi, G. Ladas, J. Popenda, J. Rakowski(Gordon and Breach Pub., Amsterdam, 2000), pp. 227-236.
\bibitem{krause2} R. Hegselmann and U. Krause, J. Artif. Soc. Soc. Simul {\bf 5}, 2 (2002).
\bibitem{fortunato} F. Fortunato, Int. J. Mod. Phys. C {\bf 16}, 259 (2005).
\bibitem{slanina} F. Slanina, Eur. Phys. J. B {\bf 79}, 99 (2011).
\bibitem{granovetter1} M. Granovetter, Am. J. Sociol, {\bf 83}, 1420 (1978).
\bibitem{axelrod1} R. Axelrod, J. Conflict Res. {\bf 41}, 203 (1997).
\bibitem{urbin1} D. Urbin, J. Lorenz, and H. Herzberg, J. Artif. Soc. Soc. Simul {\bf 11}, 4 (2008).
\bibitem{laguna1} M. F. Laguna, G. Abramson, and D. H. Zanette, Complexity, {\bf 9}, 31 (2004).
\bibitem{porfiri1}M. Porfiri, E. M. Bolt, and D. J. Stilwell, Eur. Phys. J. D {\bf 57}, 481 (2007).
\bibitem{redner1} E. Ben-Naim, P. L. Krapivsky, and S. Redner, Physica D {\bf 183}, 190 (2003).
\bibitem{pineda1} M. Pineda, R. Toral, and  E. Hern\'andez-Garc\'{\i}a, J. Stat. Mech. {\bf P08001} (2009).
\bibitem{pineda2} M. Pineda, R. Toral, and  E. Hern\'andez-Garc\'{\i}a, Eur. Phys. J. D {\bf 62}, 109 (2011).
\bibitem{jensen1} S. Grauwin and P. Jensen, Phys. Rev. E {\bf 85}, 066113 (2012).
\bibitem{carletti1} T. Carletti, D. Fanelli, A. Guarino, F. Bagnoli, and A. Guazzini, Eur. Phys. J. B {\bf 64}, 285 (2008).
\bibitem{maxi1} J. T\"or\"ok, G. Iniguez. T. Yasseri, M. S. San Miguel, K. Kaski, and J. Kertesz, Phys. Rev. Lett {\bf 110}, 088701 (2013).
\bibitem{igor} I. Douven, Logic Journal of the IGPL {\bf 18}, 323 (2010). 
\bibitem{fiedrick1} N. Friedkin, A structural theory of social influence. Cambridge, UK: Cambridge University Press, (1998).
\bibitem{fortunato1} S. Fortunato, V. Latora, A. Pluchino, and A. Rapisarda, Int. J. Mod. Phys. C {\bf 16}, 1535 (2005).
\bibitem{weron1} K. Sznajd-Weron, M. Tabiszewski, and A. M. Timpanaro, EPL {\bf 96}, 48002 (2011).
\bibitem{ben1} E. Ben-Naim, EPL {\bf 69}, 671 (2005).
\end{thebibliography}
%

\end{document}